\documentclass[journal]{aa}

\usepackage{graphicx}
\usepackage{amsmath}
\usepackage{amssymb}
\usepackage{verbatim}
\usepackage{array}
\usepackage{txfonts}
\usepackage{natbib}
\usepackage[switch]{lineno}
\usepackage{multirow}
\usepackage{float}
\usepackage{color}
\usepackage{caption}
\usepackage{xspace}  
\usepackage[acronym,nomain]{glossaries}

\bibpunct{(}{)}{;}{a}{}{,}

\newcommand{\celltspace}{\rule{0pt}{2.4ex}}
\newcommand{\cellbspace}{\rule[-1.1ex]{0pt}{0pt}}

\newcommand{\dunit}{\,cm$^{2}$\,s$^{-1}$\xspace}

\newcommand{\kms}{\,km\,s$^{-1}$\xspace}
\newcommand{\punit}{\,erg\,s$^{-1}$\xspace}

\newcommand{\eunit}{\,erg\xspace}

\newcommand{\nunit}{\,H\,cm$^{-3}$\xspace}

\newcommand{\msol}{\,M$_{\odot}$\xspace}

\newcommand{\gev}{\,GeV\xspace}
\newcommand{\tev}{\,TeV\xspace}
\newcommand{\pev}{\,PeV\xspace}
\newcommand{\gevbyc}{\,GeV/\textit{c}\xspace}
\newcommand{\tevbyc}{\,TeV/\textit{c}\xspace}
\newcommand{\mug}{\,$\mu$G\xspace}
\newcommand{\yr}{\,yr\xspace}
\newcommand{\kyr}{\,kyr\xspace}
\newcommand{\pc}{\,pc\xspace}
\newcommand{\kpc}{\,kpc\xspace}
\def\deg{\ensuremath{^\circ}\xspace}

\newacronym{ccsne}{ccSNe}{core-collapse supernovae}
\newacronym{cd}{CD}{contact discontinuity}
\newacronym{cmb}{CMB}{Cosmic Microwave Background}
\newacronym{cta}{CTA}{Cherenkov Telescope Array}
\newacronym{cr}{CR}{cosmic-ray}
\newacronym{crs}{CRs}{cosmic rays}
\newacronym{gc}{GC}{Galactic Center}
\newacronym{gps}{GPS}{Galactic Plane Survey}
\newacronym{fs}{FS}{forward shock}
\newacronym{he}{HE}{High Energy}
\newacronym{hgps}{HGPS}{H.E.S.S. Galactic plane survey}
\newacronym{iact}{IACT}{Imaging Atmospheric Cherenkov Telescope}
\newacronym{ic}{IC}{Inverse Compton}
\newacronym{ism}{ISM}{interstellar medium}
\newacronym{isrf}{ISRF}{interstellar radiation field}
\newacronym{lmc}{LMC}{Large Magellanic Cloud}
\newacronym{mw}{MW}{Milky Way}
\newacronym{psr}{PSR}{pulsar}
\newacronym[plural=PWNe,firstplural=pulsar wind nebulae (PWNe)]{pwn}{PWN}{pulsar wind nebula}
\newacronym{rs}{RS}{reverse shock}
\newacronym{sdr}{SDR}{suppressed diffusion region}
\newacronym{sn}{SN}{supernova}
\newacronym{sne}{SNe}{supernovae}
\newacronym{spp}{SPP}{SNR-PSR-PWN}
\newacronym{ts}{TS}{termination shock}
\newacronym[plural=SNRs,firstplural=supernova remnants (SNRs)]{snr}{SNR}{supernova remnant}
\newacronym{uhe}{UHE}{ultra-high-energy}
\newacronym{vhe}{VHE}{very-high-energy}

\defcitealias{Gelfand:2009}{GSZ09}
\defcitealias{Truelove:1999}{TM99}

\begin{document}

\title{Extended gamma-ray emission\\ from particle escape in pulsar wind nebulae}
\subtitle{Application to HESS J1809$-$193 and HESS J1825$-$137}
\titlerunning{Extended gamma-ray emission from particle escape in pulsar wind nebulae}
\author{Pierrick Martin\inst{\ref{irap}}\fnmsep\thanks{pierrick.martin@irap.omp.eu}
	    Louis de Guillebon\inst{\ref{irap},\ref{isae}} \and 
            Eliot Collard\inst{\ref{irap}} \and
            In\`es Mertz\inst{\ref{irap},\ref{isae}} \and
            Lars Mohrmann\inst{\ref{mpik}} \and
            Giacomo Principe\inst{\ref{infn},\ref{inaf},\ref{unitri}} \and
            Marianne Lemoine-Goumard\inst{\ref{lp2i}} \and
            Alexandre Marcowith\inst{\ref{lupm}} \and
            R\'egis Terrier\inst{\ref{apc}} \and
            Miroslav D. Filipovi\'c\inst{\ref{unisy}}}
\authorrunning{Pierrick Martin et al.}
\institute{
IRAP, Universit\'e de Toulouse, CNRS, CNES, F-31028 Toulouse, France \label{irap} \and
Institut sup\'erieur de l'a\'eronautique et de l'espace, Universit\'e de Toulouse, F-31400 Toulouse, France \label{isae} \and
LUPM, Universit\'e de Montpellier, CNRS/IN2P3, CC72, Place Eug\`ene Bataillon, F-34095 Montpellier Cedex 5, France \label{lupm} \and
Max-Planck-Institut f\"ur Kernphysik, P.O. Box 103980, D 69029 Heidelberg, Germany \label{mpik} \and
Universit\'e Bordeaux, CNRS, LP2I Bordeaux, UMR 5797, F-33170 Gradignan, France \label{lp2i} \and
Istituto Nazionale di Fisica Nucleare, Sezione di Trieste, 34127 Trieste, Italy \label{infn} \and
Dipartimento di Fisica, Universit\`a di Trieste, I-34127 Trieste, Italy \label{unitri} \and
INAF, Istituto di Radioastronomia, I-40129 Bologna, Italy \label{inaf} \and
Universit\'e de Paris, CNRS, Astroparticule et Cosmologie, 75013 Paris, France \label{apc} \and
Western Sydney University, Locked Bag 1797, Penrith South DC, NSW 2751, Australia \label{unisy}
}

\date{Received ... / Accepted ...}
\abstract
{There is growing evidence from gamma-ray observations at high and very high energies that particle escape is a key aspect shaping the morphological properties of pulsar wind nebulae (PWNe) at various evolutionary stages.}
{We aim to provide a simple model for the gamma-ray emission from these objects including the transport of particles across the different components of the system. We applied it to sources HESS J1809$-$193 and HESS J1825$-$137.}
{We developed a multi-zone framework applicable to dynamically young PWNe, taking into account the diffusive escape of relativistic electron-positron pairs out of the nebula into the parent supernova remnant (SNR) and their confinement downstream of the magnetic barrier of the forward shock until an eventual release into the surrounding interstellar medium (ISM).}
{For a wide range of turbulence properties in the nebula, the GeV-TeV inverse-Compton radiation from pairs that escaped into the remnant can be a significant if not dominant contribution to the emission from the system. It may dominate the pion-decay radiation from cosmic rays accelerated at the forward shock and advected downstream of it. In the TeV-PeV range, the contribution from particles escaped into the ISM can exceed by far that of the SNR+PWN components. Applied to HESS J1809$-$193 and HESS J1825$-$137, we found that spatially extended GeV-TeV emission components can be accounted for mostly from particles escaped into the ISM, while morphologically more compact components above $50-100$\tev are ascribed to the PWNe. In these two cases, the model suggests high turbulence in the nebula and a forward shock accelerating cosmic rays up to $\sim100$\tev at most.}
{The model provides the temporal and spectral properties of the flux of particles originally energized by the pulsar wind and ultimately released in the ISM. It can be used to constrain the transport of particles in the vicinity of pulsar-PWN-SNR systems from broadband gamma-ray observations, or in studies of the contribution of pulsar-related systems to the local positron flux.}

\maketitle

\section{Introduction}
\label{intro}

Our exploration of the sky at the highest photon energies has benefited from major advances over the past $10-15$\yr, including: (i) the extension of the accessible energy range beyond a few tens of TeV to a PeV and above \citep{Amenomori:2019,Abeysekara:2020,Cao:2021}, mostly thanks to facilities like the Tibet AS$\gamma$ experiment, the High-Altitude Water Cherenkov Observatory (HAWC), and the Large High-Altitude Air Shower Observatory (LHAASO); (ii) the investigation of gamma-ray emission on larger angular scales \citep{Abramowski:2014,Amenomori:2021,Cao:2024}, thanks to the rise of specific detection techniques and to progress in data analysis approaches \citep{Knodlseder:2019,Mohrmann:2019,Abdalla:2021}; (iii) the routine release of consistent high-level data sets covering significant portions of the sky, especially the Galactic Plane \citep{Abdalla:2018a,Albert:2020,Cao:2024}. These developments provide a broader view on the cosmic-ray phenomenon and should help us in connecting the acceleration of particles in very localized astrophysical sites to their release in the vicinity of sources and the subsequent merging into a large-scale galactic population. 

A number of recent observations seem to reveal that pulsars hold a predominant role in shaping the \gls{vhe} and \gls{uhe} appearance of our Galaxy. A significant number of gamma-ray sources above $50-100$\gev are indeed either counterparts to known pulsar-powered objects, or positionally coincident with relatively strong pulsars \citep{Abdalla:2018b,Albert:2020,Albert:2021,Cao:2021}. The observed emission is fully compatible with a pulsar origin in terms of spectral shape, maximum detected energy, and flux level \citep{Sudoh:2021,Breuhaus:2021,Breuhaus:2022,DeOnaWilhelmi:2022}, and population synthesis efforts confirm that the majority of currently known TeV sources are powered by pulsars \citep{Fiori:2022,Martin:2022b}. A reliable exploitation of the growing body of gamma-ray observations therefore necessitates a solid understanding of pulsar-related emission, all the more so that deep and extensive surveys with next generation instruments like the \gls{cta} will be undertaken soon \citep{Acharyya:2023,Abe:2023} and are expected to significantly increase the number of known pulsar-powered gamma-ray sources.

Pulsar-powered objects in the \gls{vhe}/\gls{uhe} range are primarily \glspl{pwn}, sources typically $3-30$\pc in physical size involving pulsars with characteristic ages $1-100$\kyr \citep{Abdalla:2018b}. The recent discovery of TeV halos or pulsar halos \citep{Linden:2017,Abeysekara:2017b,Linden:2018} seems to extend the contribution of pulsars in the \gls{vhe}/\gls{uhe} sky to larger physical extents (possibly $>30$\pc) and older systems ($>100$\kyr pulsars). It is not clear how this new kind of object relate to \glspl{pwn} in terms of evolutionary path \citep[see the reviews by][]{Fang:2022,LopezCoto:2022}, and how frequently the phenomenon occurs \citep{Giacinti:2020,Martin:2022}. Part of the problem stems from the difficulty to characterize observationally the physical boundaries of the system and the medium that radiating particles are located in; depending on whether this is undisturbed \gls{ism}, stellar ejecta in the \gls{snr}, or shocked pulsar wind, the physical interpretation differs. Pulsar halos are a window on late stages of (some) pulsar-related systems and non-thermal particle transport in the vicinity of accelerators, and our inability to apprehend them reflects our limited understanding of both these topics. 

As we gain access to lower surface brightness emission over larger angular scales, there is growing evidence that particle escape or leakage across the various components of the system is a key aspect in accounting for the properties of both dynamically young and evolved pulsar-related systems. This applies for instance to HESS J1825-137 \citep{Abdalla:2019,Principe:2020}, HESS J1809-193 \citep{Aharonian:2023}, HESS 1813-178 \citep{Aharonian:2024b}, Vela X \citep{Hinton:2011}, or Geminga and pulsar B0656+14 \citep{Abeysekara:2017b}. 

The goal of this paper is therefore to propose a model for the gamma-ray emission from the whole system harbouring a \gls{pwn}, taking into account the diffusion of particles out of the \gls{pwn} into the \gls{snr} and subsequently to the surrounding \gls{ism}. Based on previous modelling efforts \citep[][hereafter GSZ09]{Gelfand:2009}, we implemented a one-zone model for the dynamics and radiation from the \gls{pwn} including a prescription for the properties of magnetic turbulence and the corresponding diffusive escape. The originality of our work is to provide a description for the subsequent fate and emission of escaping particles, namely their trapping in the remnant for some time followed by their eventual release in the vicinity of the source. The model framework is introduced in Sect. \ref{mod}, we examine its typical output in Sect. \ref{res}, and apply it to the case of HESS J1809$-$193 and HESS J1825$-$137 in Sect. \ref{app}. We summarise our work in Sect. \ref{conclu}.

\section{Modelling framework}
\label{mod}

The model framework presented here extends that introduced in \citetalias{Gelfand:2009}, of which we provide a detailed description including the most relevant quantities and formulae in Appendix \ref{app:mod}. A first additional feature is the possibility of diffusive escape from the nebula, computed under the assumption of particle scattering in a fully-developed Alfvenic turbulence. A second additional feature is the description of the temporary trapping of escaped particles in the remnant, followed by their eventual escape into the interstellar medium. In this section, we first provide a brief qualitative description of the main ideas behind the original framework of \citetalias{Gelfand:2009}, and then we introduce the assumptions underlying our model.

\subsection{Main features of the original model}
\label{mod_org}

The model initially presented in \citetalias{Gelfand:2009} describes a fast relativistic outflow emanating from the pulsar, whose size is neglected, and its interaction with the expanding stellar ejecta resulting from the parent supernova explosion. The model assumes spherical symmetry of the system and we will henceforth use $r$ to denote the position from the centre of the system and $t$ the time since supernova explosion and simultaneous appearance of the pulsar. We neglect any motion of the pulsar with respect to the system (but the original model includes the possibility of a non-zero velocity of the pulsar and its escape from the nebula).

The stellar ejecta interact with a cold circumstellar medium and rapidly take a characteristic structure, in which the expanding remnant drives a \gls{fs} that propagates in the ambient medium and has position $R_{\rm SNR}(t)$, while a \gls{rs} propagates back in the ejecta and has position $R_{\rm RS}(t)$. The description of this double-shock dynamics is based on analytical formulae that hold only in the non-radiative stage, when energy losses from thermal radiation of the system are negligible.

Meanwhile, the central pulsar is assumed to spin down as a result of magnetic dipole radiation, which powers a relativistic outflow, the pulsar wind. At sufficiently large distance from the pulsar, the wind is mostly composed of a toroidal magnetic field and electron-positron pairs. Upon interaction with the surrounding medium, the wind is halted at a \gls{ts}, where particles are efficiently accelerated up to very high energies. The acceleration mechanism is not yet fully elucidated and may involve several sites and processes \citep[diffusive shock acceleration at the TS, turbulent acceleration downstream of it, etc; see][]{Bucciantini:2011,Amato:2020}, but observations indicate that it drains a large fraction of the wind kinetic energy, of the order of several tens of percent \citep{Zhang:2008,Bucciantini:2011,Torres:2014}.

The \gls{pwn} is the volume excavated by the shocked pulsar wind as it expands outwards, treated as a single zone in which quantities are assumed to be uniform (including energy injection). It is bounded on the inner side at a radius $R_{\rm TS}(t)$ by the \gls{ts}, and on the outer side at a radius $R_{\rm PWN}(t)$ by a thin shell of swept-up ejecta material. Energy is injected in the nebula in the form of non-thermal pairs and magnetic field. The original model assumes magnetic flux conservation to determine the evolution of magnetic energy as the volume of the nebula changes, while the internal energy of non-thermal particles evolves as a result of adiabatic and radiation losses.

The bounding shell mass is made of ejecta material collected when the shell velocity exceeds the ejecta velocity immediately ahead of it. Its dynamics is controlled by its inertia and the pressure difference between the nebula on the inner side, and the stellar ejecta on the outer side. At early times, the innermost unshocked stellar ejecta are assumed to have a negligible pressure because of their fast expansion and rapid cooling. Later on, the reverse shock propagating back through the ejecta eventually reaches the nebula, heats and compresses the material ahead of the bounding shell such that the pressure immediately outside the nebula becomes non-negligible and possibly well in excess of that in the nebula. This may lead to a strong compression of the nebula, initiating several cycles of compression and re-expansion referred to as reverberation.

The spectral distribution of particles at any given time depends on the assumed injection properties (pulsar spin-down history and initial particle spectrum), the magnetic and radiation fields in which they are immersed (which sets the magnitude of radiative losses), and the dynamics of the nebula (which drives adiabatic energy losses or gains). It is used to compute the broadband radiation from the nebula, consisting of synchrotron emission in the nebular magnetic field and inverse-Compton scattering in the ambient photon fields.

\subsection{Additional developments}
\label{mod_add}

\subsubsection{Particle escape from the nebula}
\label{mod_add_esc}

We modified the energy injection equation to include a term for turbulence, not part of the original model of \citetalias{Gelfand:2009}:
\begin{align}
\dot{E}_{\rm inj} (t) &= \dot{E}_{\rm inj,B} (t) + \dot{E}_{\rm inj,e} (t) + \dot{E}_{\rm inj,T} (t) = \dot{E}_{\rm PSR} (t) \, , \\
&= \eta_{\rm B} (t) \dot{E}_{\rm PSR} (t) + \eta_{\rm e} (t) \dot{E}_{\rm PSR} (t) + \eta_{\rm T} (t) \dot{E}_{\rm PSR} (t) \, .
\end{align}
Term $\dot{E}_{\rm inj,e}$ (and corresponding $\eta_{\rm e}$) describes the injection of relativistic pairs in the nebula. Term $\dot{E}_{\rm inj,B}$ (and corresponding $\eta_{\rm B}$) describes a large-scale regular magnetic field component, similar to the original model, while term $\dot{E}_{\rm inj,T}$ (and corresponding $\eta_{\rm T}$) denotes a turbulent magnetic field component. 

Turbulence is assumed to be composed of Alfv\'en waves so the injected energy consists for one half of kinetic energy of the turbulent fluid, and for the other half of its magnetic energy. For simplicity and consistency, the evolution with time and volume of turbulent energy and pressure is taken to be similar to what was assumed in \citetalias{Gelfand:2009} for the large-scale magnetic field term.

The turbulence spectrum is ideally obtained by solving a complete wave transport equation for a given phenomenology (e.g. Kolmogorov or Kraichnan). In our context, turbulence is expected to be injected at a spatial scale which is a fraction of the \gls{ts} radius. Turbulence cascading as a result of non-linear wave-wave interactions, which can be described as diffusion in wavenumber space \citep{Miller:1995}, transports turbulent energy to larger wavenumbers (smaller physical scales), while volume expansion pushes turbulence towards smaller wavenumbers (larger physical scales). 

For typical parameters of the problem, the Alfv\'en speed is quite high, up to a few tens of percent of the speed of light, such that the development of the cascade occurs on time scales that are much shorter than the dynamical time scale. We therefore considered a simplified description of turbulence, avoiding the solving of a complete transport equation, and assumed that the magnetic turbulence spectrum is fully developed at each time, from a maximum scale that is a fraction $\kappa_T \leq 1$ of the radius of the nebula down to an arbitrarily small scale. In practice:
\begin{align}
\label{eq:wk}
&w(k,t) = \frac{1}{3} \frac{E_{\rm PWN,T}(t)}{V_{\rm PWN}(t)} \frac{1}{ \left( k_{\rm inj}^{-2/3} - k_{\rm cut}^{-2/3} \right)} k^{-5/3} \, ,\\
& k_{\rm inj}(t) = \frac{2 \pi}{\kappa_T R_{\rm PWN}(t)} \textrm{ and } k_{\rm cut}(t) \gg k_{\rm inj}(t) \, .
\end{align}
A justification for the assumption of fully-developed turbulence at all times is provided in Appendix \ref{app:time}. We do not describe the energization of particles by magnetic reconnection in the turbulent flow, or stochastic scattering off turbulent fluctuations \citep{Luo:2020,Lu:2023}. These processes were considered in the context of the origin of the radio-emitting particle population, and in relation to the question of the confinement of the nebula. An implicit hypothesis in our model framework is that the conversion of ordered toroidal field into magnetic turbulence and the dissipation of magnetic energy occur on sufficiently small temporal and spatial scales downstream of the \gls{ts}, and that the assumed turbulent state and particle injection spectrum in our model may be the result of these processes.

Electron-positron pairs are assumed to experience advection in the nebula, from the termination shock to the bounding shell. Advection is assumed to occur following the prescription used in \citet{Zhu:2023}, which involves a $1/r$ velocity profile down to a minimal value that is the velocity of the thin bounding shell, while diffusion results from resonant interactions with magnetic turbulence on the scale of the Larmor radius of the particle, which is described by the coefficient:
\begin{align}
\label{eq:diffcoef}
& D_{\rm PWN} (p,t) = \frac{1}{3} \frac{c r_{\rm L} (p,t)}{k_{\rm res} W(k_{\rm res},t)} \, , \\
& W(k,t) = \frac{w(k,t)}{\frac{E_{\rm PWN,B}(t) + E_{\rm PWN,T}(t)/2}{V_{\rm PWN}(t)}} \, ,
\end{align}
where $p$ is the momentum of the particle. We used the approximation introduced in \citet[][their Eq. 5]{Ptuskin:2003}, valid for an arbitrary turbulence level, in which the Larmor radius $r_{\rm L}$ is taken in the total ordered+turbulent magnetic field, $k_{\rm res} = 2 \pi / r_{\rm L}$ is the corresponding wavenumber at the resonant scale, and $W(k)$ is the turbulence spectrum normalized to the ordered+turbulent magnetic field energy density.

The spectral distribution of the escaping flux at each time, $Q_{\rm esc,PWN} (E,t)$, is computed from the spectral distribution of particles in the nebula $N_{\rm e,PWN}(E,t)$ and the escape time scale $\tau_{\rm esc,PWN} (E,t)$:
\begin{align}
\label{eq:escflux}
&Q_{\rm esc,PWN} (E,t) = -\frac{N_{\rm e,PWN}(E,t)}{\tau_{\rm esc,PWN} (E,t)} \, , \\
&\tau_{\rm esc,PWN} (E,t) = \left( \frac{1}{\tau_{\rm diff,PWN}} + \frac{1}{\tau_{\rm adv,PWN}} \right)^{-1} \, ,
\end{align}
where the escape time scale combines the diffusion and advection crossing times $\tau_{\rm diff,PWN}$ and $\tau_{\rm adv,PWN}$:
\begin{align}
\label{eq:esctau}
&\tau_{\rm diff,PWN} (E,t) = \frac{R^2_{\rm PWN}(t)}{6 D_{\rm PWN}(E,t)} \, , \\
&\tau_{\rm adv,PWN} (t) = \frac{\left( R^2_{\rm PWN}(t) - R^2_{\rm TS}(t) \right)}{2 R_{\rm PWN}(t) v_{\rm PWN}(t)} \, ,
\end{align}
where the diffusion coefficient is expressed as a function of kinetic energy $E$ rather than momentum $p$ as in Eq. \ref{eq:diffcoef}. We assume here three-dimensional diffusion, although we did not make any explicit assumption on the topology of the ordered field. The escape flux term is included on the right-hand side of Eq. \ref{eq:ne}. 

This implementation essentially describes the nebula as a magnetized turbulent bubble that holds particles only until they reach its outskirts, and implies that the surrounding bounding shell plays no role in confining them. This assumption can be justified from radio polarimetric observations of the Crab showing that the magnetic field in the nebula evolves from mostly toroidal close to the termination shock to predominantly radial at larger distances \citep{Bietenholz:1990}. Particles can then slide outwards along the field lines, which could explain the limited spectral index variations measured in X-rays along filaments in the outer regions \citep{Seward:2006}.

\begin{figure}[!t]
\begin{center}
\includegraphics[width=0.9\columnwidth]{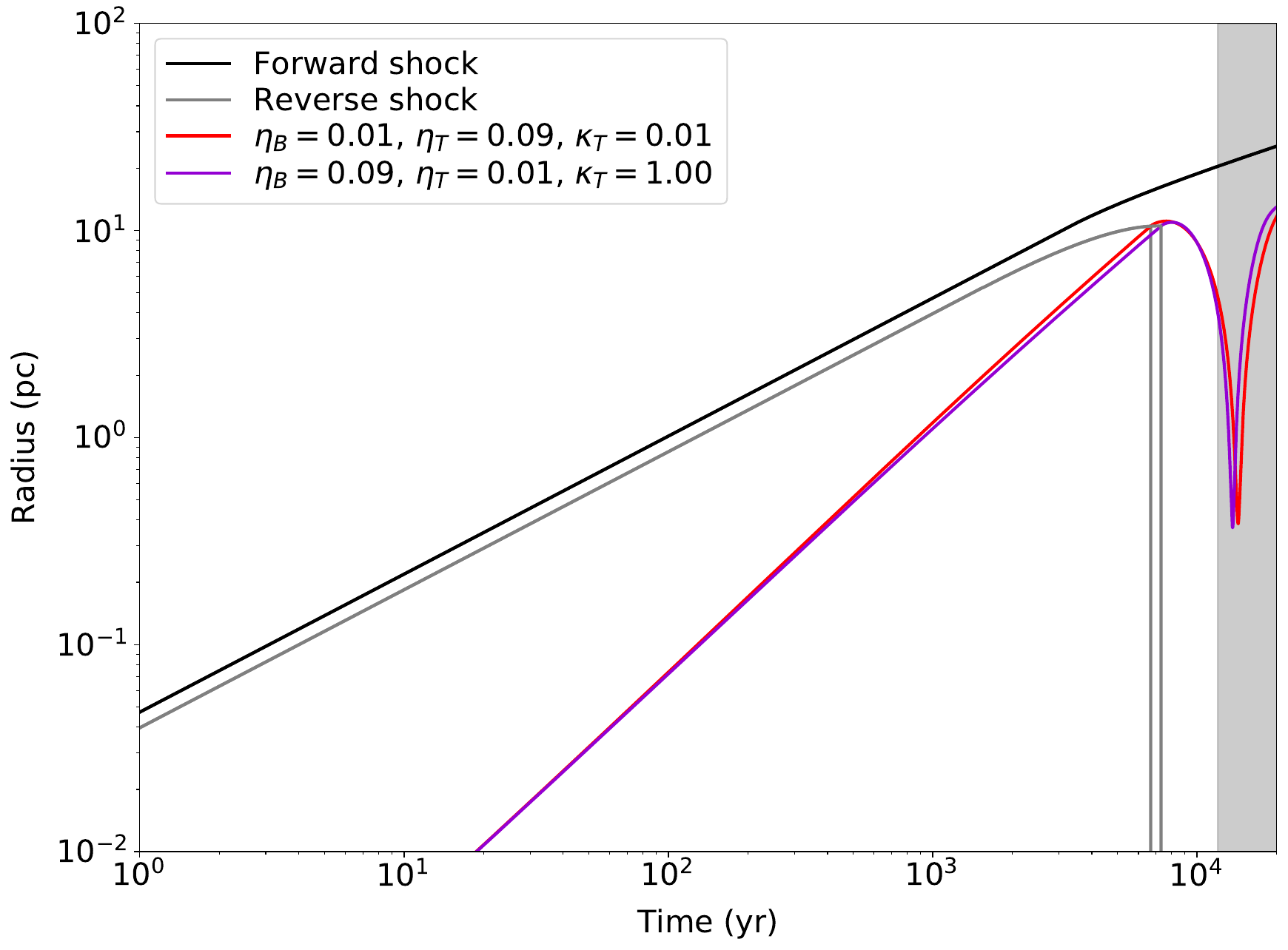}
\caption{Evolution of the size of the \gls{pwn} for the two extreme model setups. Also shown as reference are the radii for the forward and reverse shocks of the \gls{snr}. The curves for the other setups lie in between the red and purple curves. In this and subsequent figures, the gray shaded area denotes the strong compression phase that the model is not suited to describe.}
\label{fig:res:pwnradius}
\end{center}
\end{figure}

\subsubsection{Particle trapping in the remnant}
\label{mod_add_snr}

Particles escaping from the nebula enter the remnant. The transport of pairs across the remnant will depend on the actual magnetic conditions in the volume and may well be pretty complicated. There are frontiers that particles may not easily cross before reaching the outer \gls{fs}, first and foremost the \gls{pwn} bounding shell but also the ejecta-\gls{ism} contact discontinuity, both of which are expected to be turbulent because of hydrodynamical instabilities. This complexity is beyond the scope of our model and we adopted a minimalist description of particle confinement in the \gls{snr} volume, focussing on what happens at its outer edge.

Because of particle acceleration and magnetic field amplification at the \gls{fs}, that frontier acts as a magnetic barrier confining energetic particles up to a certain maximum momentum downstream of the shock. That maximum momentum evolves with time and we adopted the prescription used in \citet{Celli:2019}:
\begin{equation}
\label{eq:pmax}
p_{\rm max}(t) =  \begin{cases} p_{\rm max}^{\rm ST} (t/t_{\rm ST}) & t \leq t_{\rm ST} \, , \\
p_{\rm max}^{\rm ST} (t/t_{\rm ST})^{-\xi} & t > t_{\rm ST} \, . \end{cases}
\end{equation}
Equation \ref{eq:pmax} implicitly encapsulates the intricate physics governing the evolution of the amplified magnetic field at the \gls{fs}. We handle the absolute maximum momentum $p_{\rm max}^{\rm ST}$ and the time dependence index $\xi$ as free parameters of the model, with typical values in the range $\sim 0.1-1$\pev and $\sim2.0-4.0$, respectively. As an example, the interpretation of gamma-ray observations of the $\gamma$-Cygni SNR in the framework of an escape model based on that prescription resulted in estimated values of $p_{\rm max}^{\rm ST}=78$\tev/c and $\xi=2.55$ \citep{Acciari:2023}.

The SNR is treated as a one-zone region hosting a specific population of energetic particles, fed by a continuous inflow of particles escaping from the PWN and progressively depleted of particles with energies above $p_{\rm max}(t)$. As the SNR evolves, the confined particles lose energy as a result of adiabatic expansion, inverse-Compton scattering, and synchrotron radiation. Particles trapped in the \gls{snr}, until their complete release into the \gls{ism} when the remnant is deep into the Sedov-Taylor stage, power an extended emission component around the \gls{pwn}, with a significant flux comparable to or well in excess of that of the nebula in some cases, as we will see below. 

The magnetic field in the remnant can have several origins: swept-up interstellar magnetic field, possibly including some amplification as a result of particle acceleration at the \gls{fs}, magnetic energy frozen in the stellar ejecta, magnetic field amplification by particles streaming out of the nebula,... These different contributions are expected to have specific spatial distributions, and in some cases to be restricted to compressed layers or filaments. For simplicity, and because of the one-zone assumption for this volume, we considered a magnetic field in the remnant similar to that in the surrounding \gls{ism}. This is obviously a lower limit case, mostly relevant to evolved systems with ages $\gtrsim 2-3$\kyr and size $\gtrsim 10-20$\pc.

\section{Model predictions}
\label{res}

\subsection{Model applicability}
\label{res_caveats}

We present below predictions for a reference pulsar-\gls{pwn}-\gls{snr} system. We caution beforehand that the application of the model framework defined in Sect. \ref{mod} should be restricted to early evolutionary stages, which we define as the time period from pulsar birth through free expansion of the nebula and up to beyond reverse-shock interaction until modest levels of compression of the \gls{pwn} are reached (less than a factor of ten). 
Indeed, it was demonstrated that analytical prescriptions for the hydrodynamical properties of the remnant such as those used in our model are inappropriate to accurately describe the evolution of the system past reverse-shock crushing \citep{Bandiera:2023b,Bandiera:2023c}. Second, reverse-shock crushing can reasonably be expected to be asymmetric, thereby reflecting gradients in the surrounding \gls{ism} density, which combined with a non-zero pulsar kick not necessarily aligned with the density gradient can give rise to a myriad of situations that our spherical model can definitely not describe properly \citep[see examples in][]{Kolb:2017}. 
In its current version, the model cannot be used to describe very evolved systems, for instance those with two detached nebulae (a relic nebula and a younger bow-shock nebula in the \gls{snr} or \gls{ism}) or TeV halos. 

We also emphasise that we do not describe the spatial transport of particles that manage to escape into the \gls{ism}. As it is, the model cannot self-consistently predict the transport conditions in that medium, so we only compute the spectral evolution of the escaped particles in time and the corresponding integrated non-thermal emission. Our purpose is to show that the \gls{ism} contribution can be significant in terms of total signal (see Sects. \ref{res_gamlum} and \ref{res_gamspec} below), such that observations should tell us something about diffusion/transport conditions close to the source.

\subsection{Model setups}
\label{res_setups}

In our reference system, the \gls{snr} component is parametrized as an ejecta of mass $M_{\rm ej}=10$\msol and initial kinetic energy $E_{\rm ej}=10^{51}$\eunit, expanding into a uniform medium with hydrogen density $n_{\rm ISM}=0.1$\nunit and mean molecular mass $\mu=1.4$. The initial ejecta density profile consists of a uniform core and a power-law envelope with index $n_{\rm ej}=9$, and the initial velocity profile is a linear growth starting from zero at the centre. Particle confinement within the remnant is defined by default by $p_{\rm max}^{\rm ST}=10^5$\gevbyc and $\xi=2.5$, but we explored variations of these parameters. 

The pulsar component has an initial spin-down power $L_{0}=5 \times 10^{38}$\punit and spin-down time scale $\tau_{0}=2000$\yr. The pulsar spins down over time with a constant braking index $n_{\rm PSR}=3$. For simplicity, we set the natal kick velocity of the pulsar to zero. At the wind termination shock, a fraction $\eta_{e}=90$\% of the wind power is converted into non-thermal electron-positron pairs, and we consider various possibilities for the splitting of the remaining 10\% into ordered magnetic field and Alfvenic turbulence (via parameters $\eta_{\rm B}$ and $\eta_{\rm T}$). Downstream of the termination shock, non-thermal particles are injected in the nebula with an energy spectrum consisting of a broken power-law with indices $\alpha_1=1.5$ and $\alpha_2=2.3$ respectively below and above a break energy $E_{\rm brk}=500$\gev, while an exponential cutoff terminates the spectrum at $E_{\rm cut}=1$\pev.

The whole system is located at a distance of 3\,kpc from us, in the direction of the Galactic centre. The interstellar radiation field density at this position is taken from the model of \citet{Popescu:2017}, and the interstellar magnetic field is assumed to have a strength of $B_{\rm ISM}=5$\mug.

We explore the effects of different prescriptions for the turbulence, hence different particle escape profiles, on the resulting non-thermal emission. We considered two families of setups: low level of magnetic turbulence ($\eta_B=0.09$  and $\eta_T=0.01$), high level of magnetic turbulence ($\eta_B=0.01$  and $\eta_T=0.09$), and for each case we tested three spatial scales for turbulence injection ($\kappa_T = 0.01,0.1,1.0$). 
We emphasise that, because turbulent energy is half kinetic and half magnetic in the adopted Alfvenic turbulence, the different families of setups correspond to different levels of magnetic energy injection: 9.5\% for the L setups, and 5.5\% for the H setups. In addition to the impact on particle escape, there will thus be some impact on synchrotron radiation losses, hence on particle spectra at the high energy end.

The different $\eta_B$, $\eta_T$, and $\kappa_T$ parameter combinations have dynamical implications, in the sense that they have consequences on the amount and form of the energy contained in the \gls{pwn}, hence on its expansion rate or ability to withstand compression. Conversely, parameters $p_{\rm max}^{\rm ST}$ and $\xi$ controlling particle escape from the \gls{snr} into the \gls{ism}
have no effect on the dynamics of the system, and they only affect the radiation from the \gls{ism} relative to that from the \gls{snr}.

\begin{figure}[!t]
\begin{center}
\includegraphics[width=0.9\columnwidth]{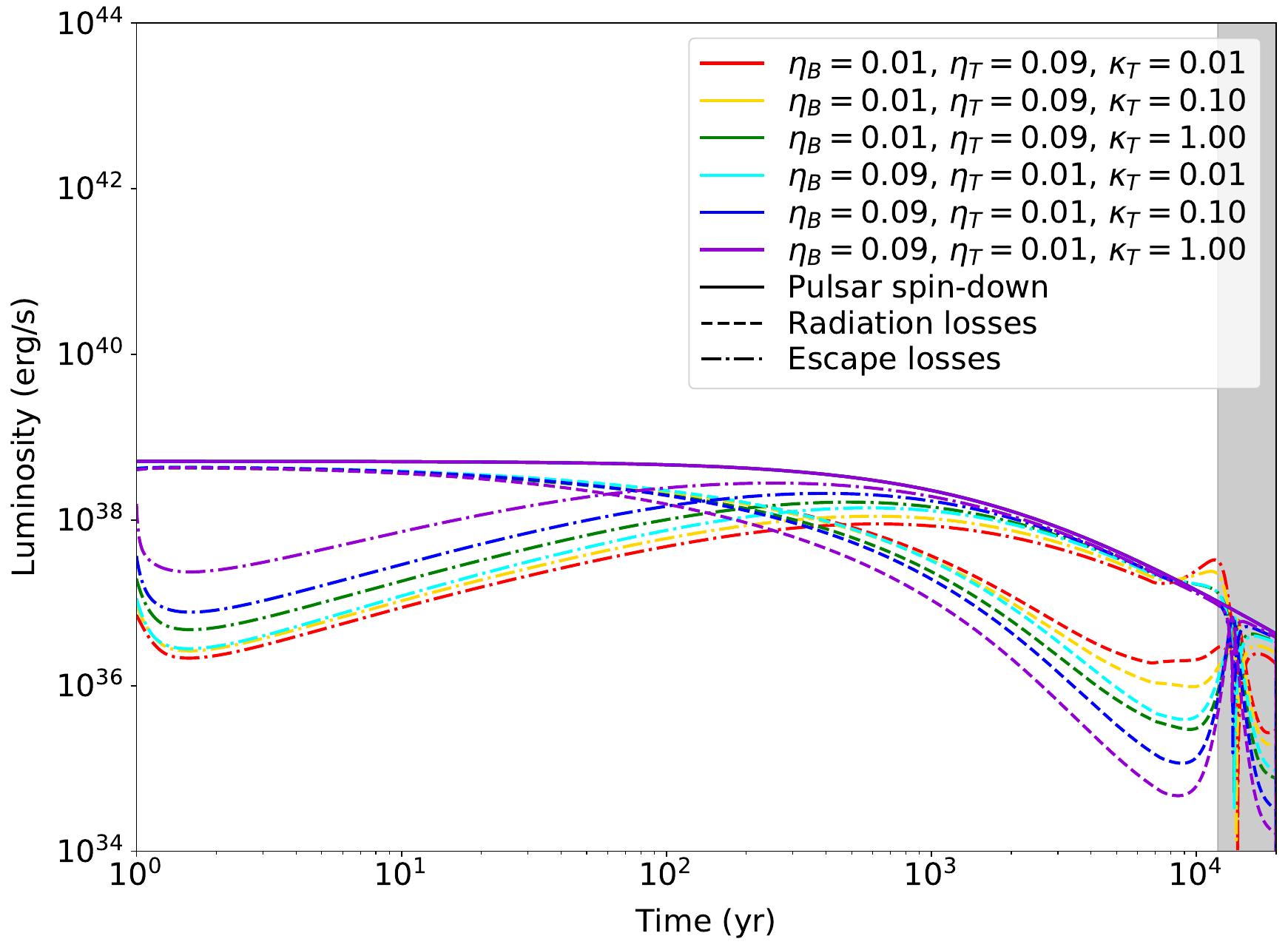}
\caption{Evolution of the radiation and escape luminosities in the \gls{pwn} volume for the different model setups. Also shown as reference is the pulsar spin-down power.}
\label{fig:res:pwnlum}
\end{center}
\end{figure}

\begin{figure}[!t]
\begin{center}
\includegraphics[width=0.9\columnwidth]{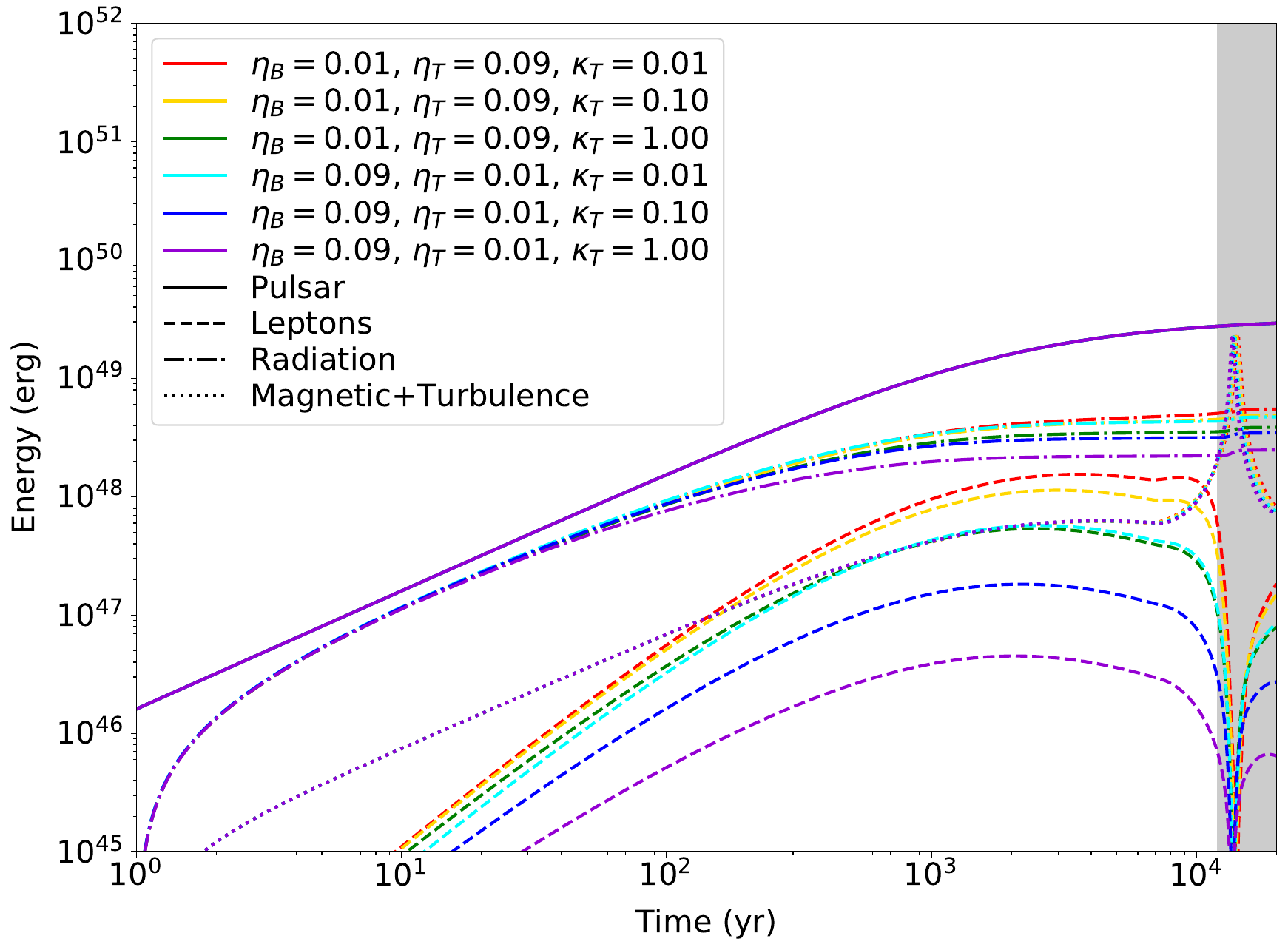}
\caption{Evolution of the energy content in the \gls{pwn} volume for the different model setups. Also shown as reference is the cumulative energy injected by the pulsar and the cumulative energy lost to radiation in the nebula.}
\label{fig:res:pwnnrj}
\end{center}
\end{figure}

\begin{figure}[!t]
\begin{center}
\includegraphics[width=0.9\columnwidth]{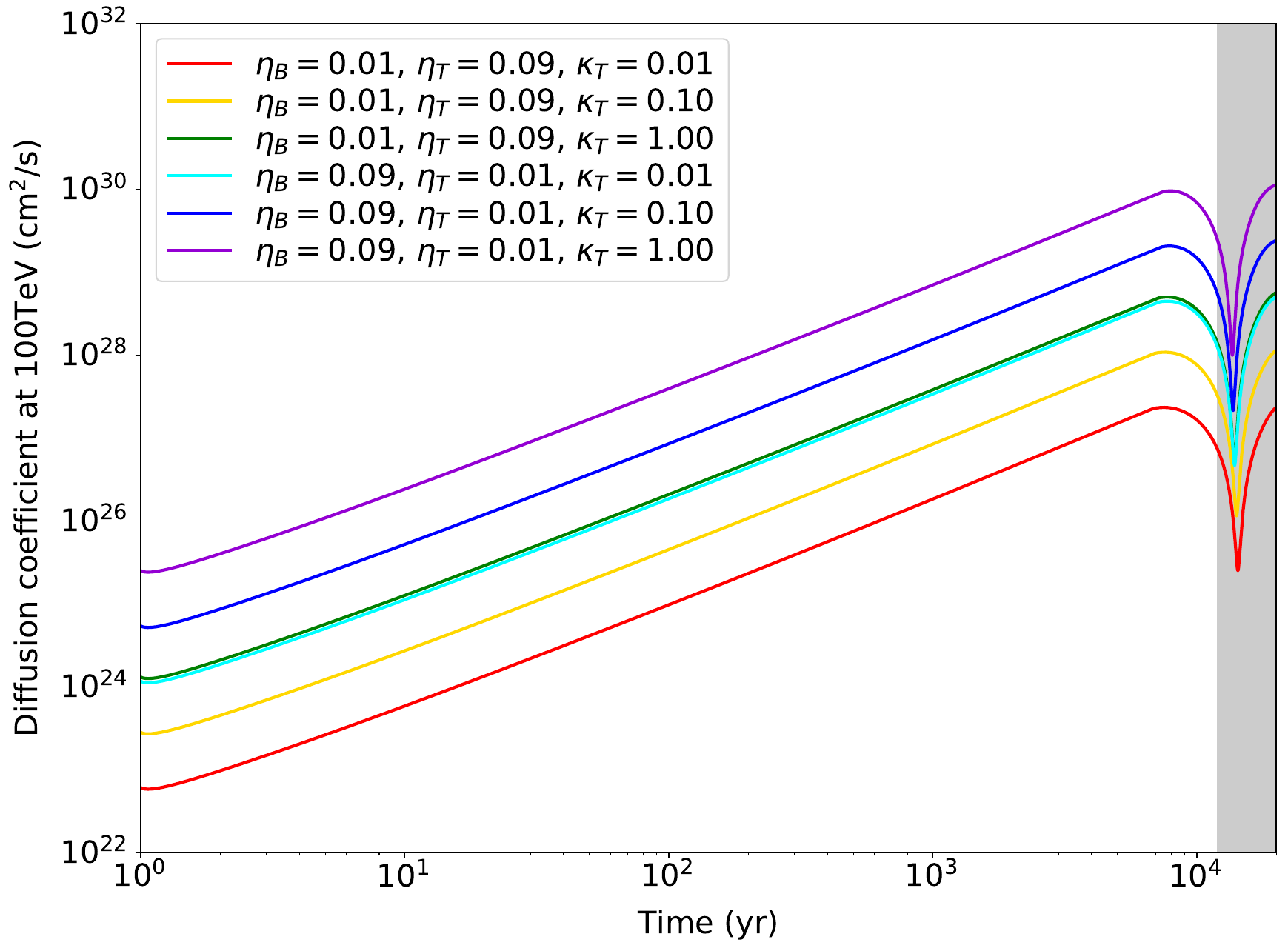}
\caption{Evolution of the spatial diffusion coefficient at 100\tev in the different model setups. For reference, the effective value of the diffusion coefficient at 100\tev for large-scale transport in the Galactic disk is on the order of $10^{30}$\dunit.}
\label{fig:res:diffcoef}
\end{center}
\end{figure}

\subsection{Energetics and dynamics}
\label{res_dyna}

\begin{figure}[!t]
\begin{center}
\includegraphics[width=0.9\columnwidth]{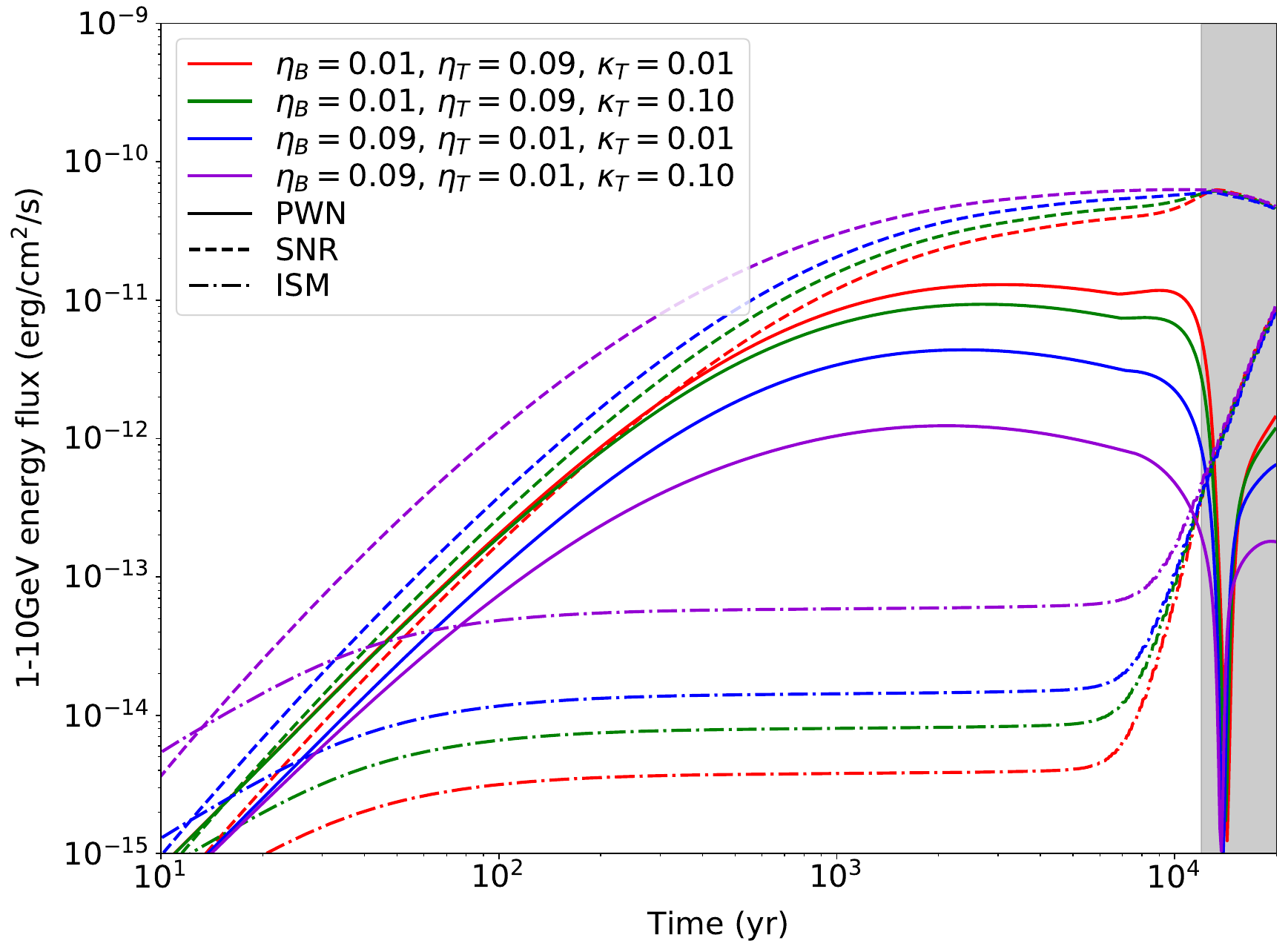}
\includegraphics[width=0.9\columnwidth]{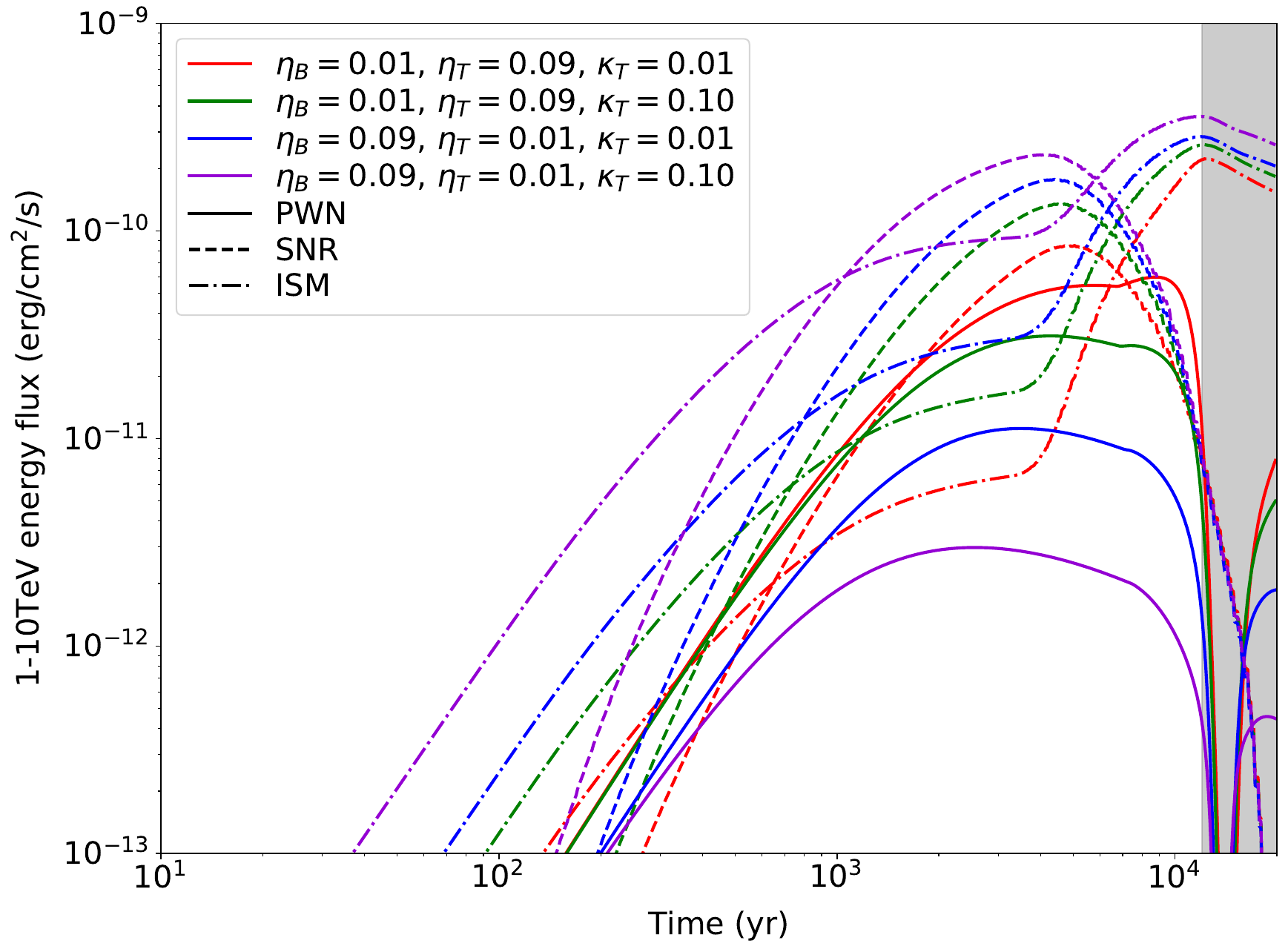}
\includegraphics[width=0.9\columnwidth]{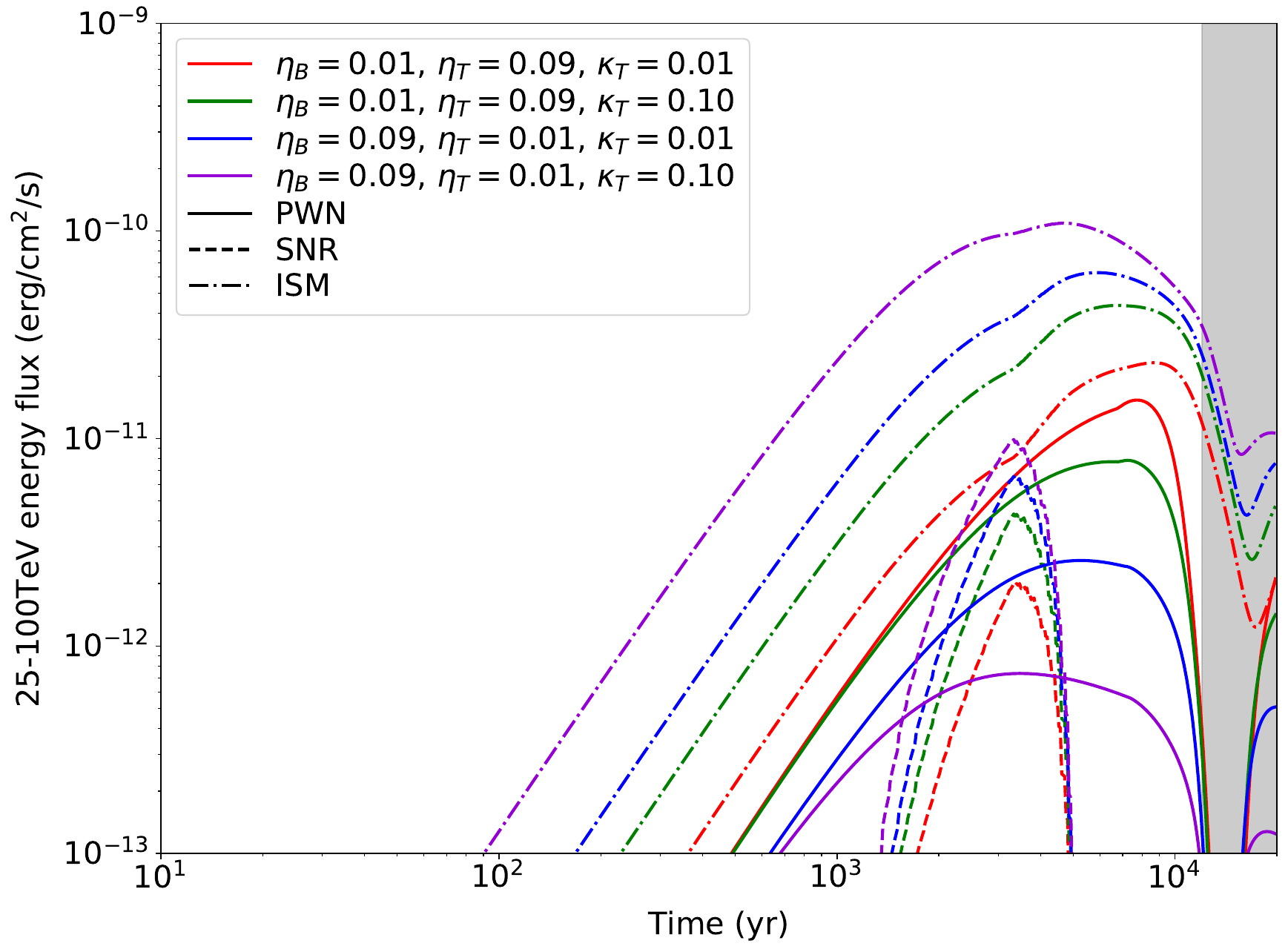}
\caption{Evolution of the gamma-ray luminosity in different bands for the \gls{pwn}, \gls{snr}, and \gls{ism} components in four model setups.}
\label{fig:res:gamlum}
\end{center}
\end{figure}

\begin{figure}[!t]
\begin{center}
\includegraphics[width=0.9\columnwidth]{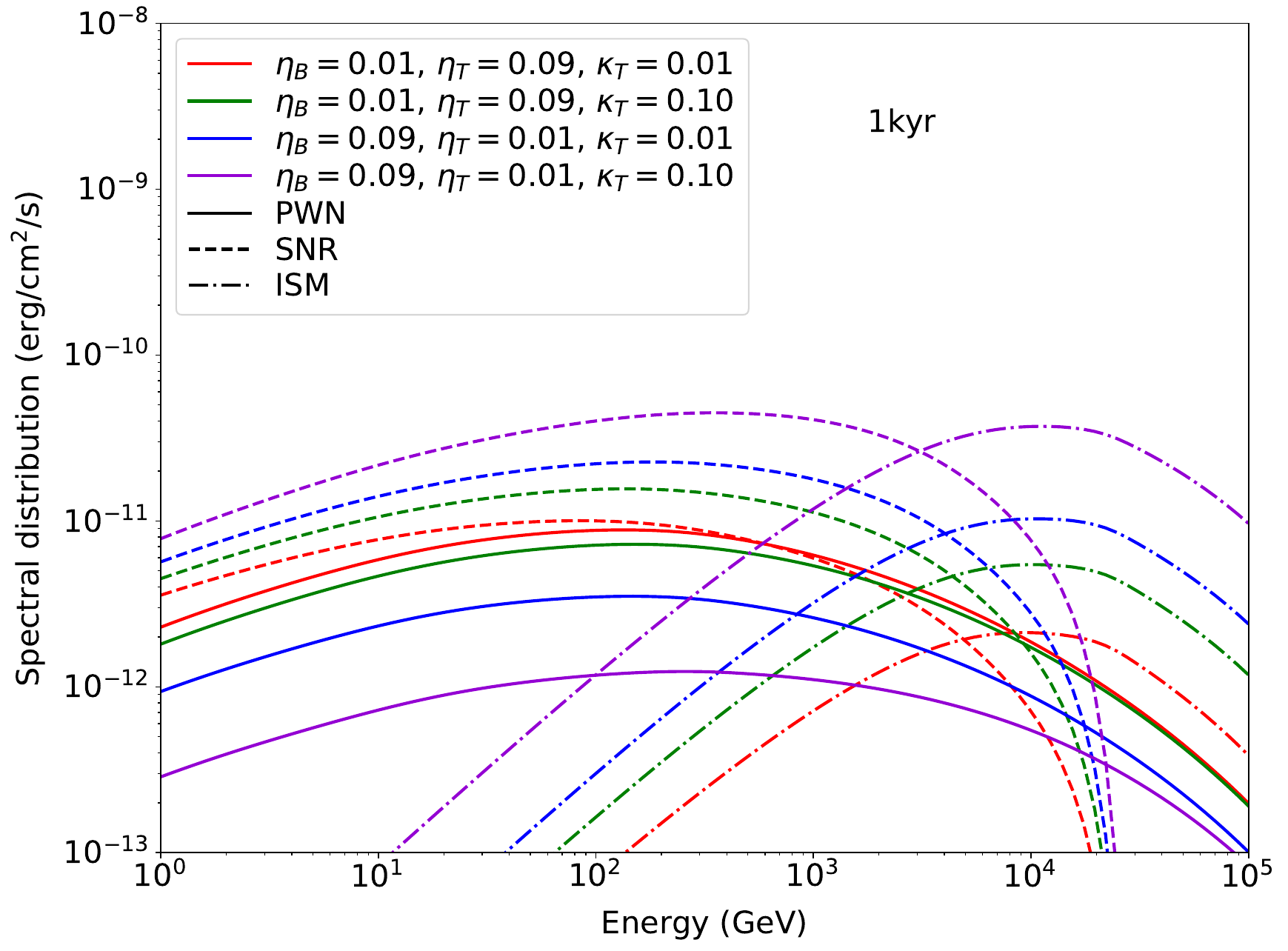}
\includegraphics[width=0.9\columnwidth]{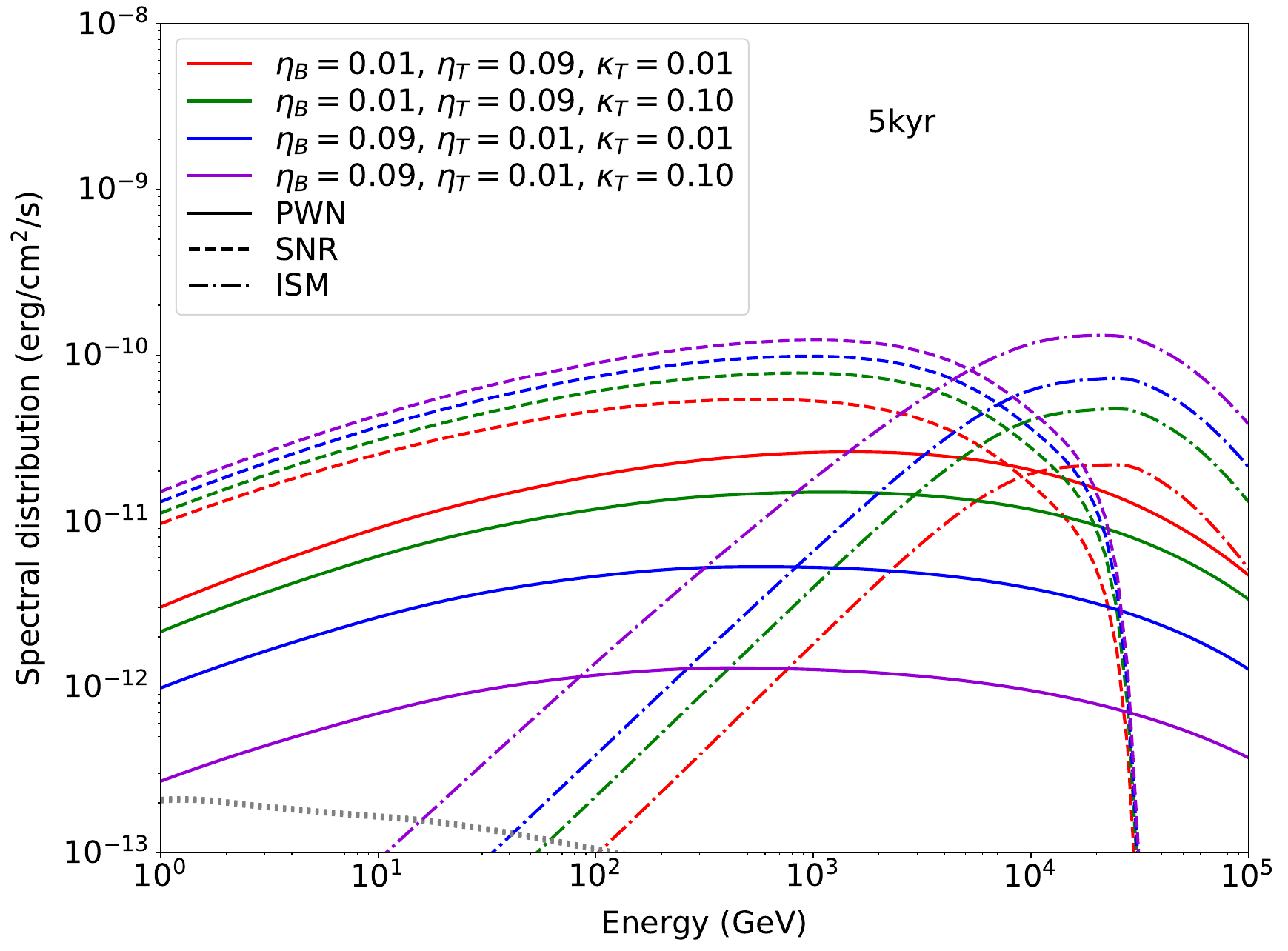}
\includegraphics[width=0.9\columnwidth]{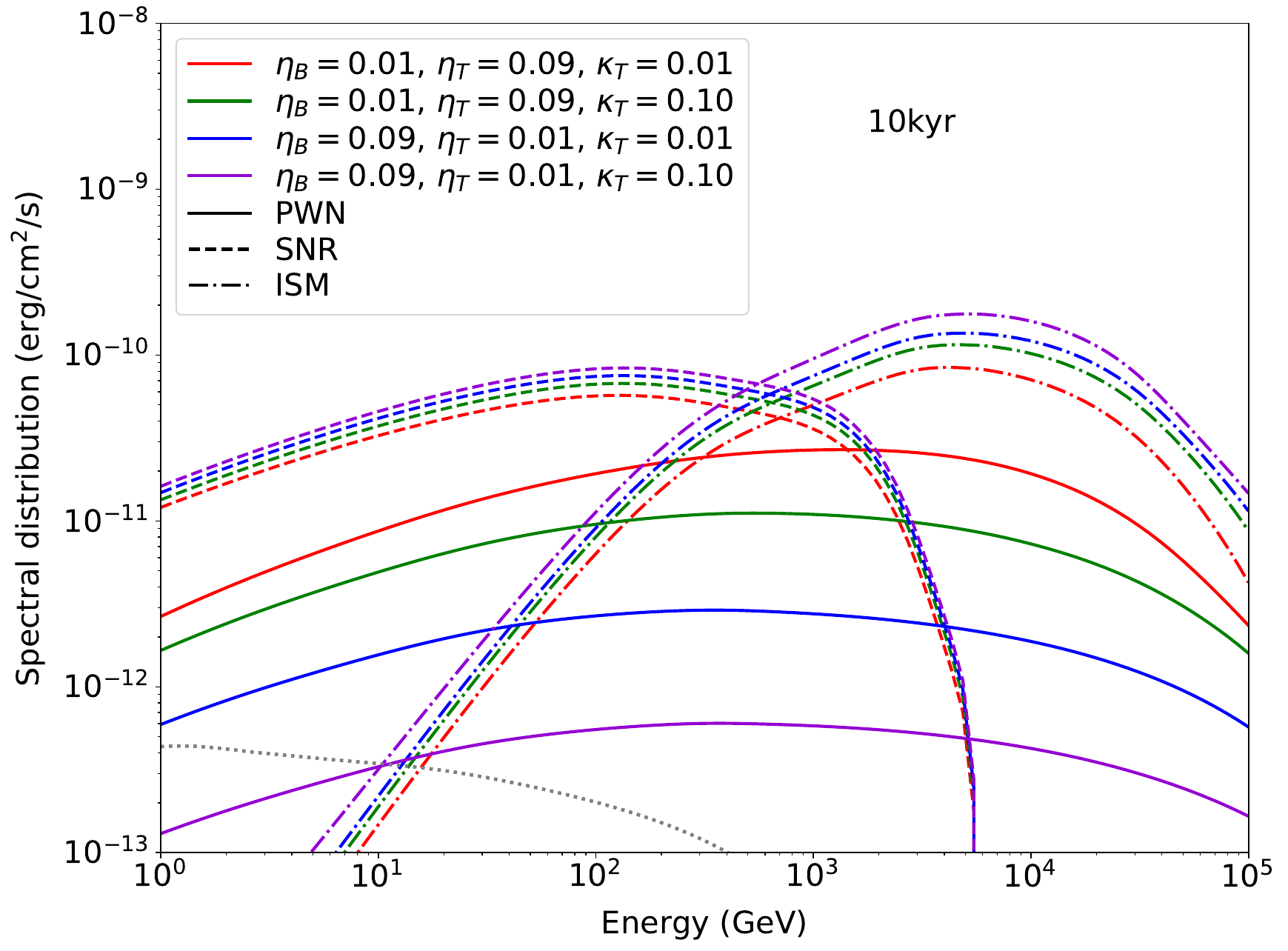}
\caption{Gamma-ray spectra of the \gls{pwn}, \gls{snr}, and \gls{ism} components in four model setups at different ages. The gray dotted line corresponds to an order-of-magnitude estimate of the pion-decay emission from cosmic rays in the remnant (see text for details).}
\label{fig:res:gamspec}
\end{center}
\end{figure}

Figure \ref{fig:res:pwnradius} displays the dynamics of the forward and reverse shocks in the remnant, and of the thin bounding shell forming the outer frontier of the nebula. The reverse shock hits the nebula at $t \simeq 7000$\yr, which triggers its compression only about one thousand years later owing to the inertia of the bounding shell.

Figure \ref{fig:res:pwnlum} displays the time evolution of particle escape compared to that of radiative losses and pulsar spin-down. One can first note the similarity of the curves in the two cases with highest turbulence (red and orange), which reflects the fact that advection is then the dominant escape process. Our model suggests a maximum in escape losses somewhat around the decrease of the pulsar spin-down power at $t \simeq \tau_{0}$. In high-turbulence cases like H1 and H10, there is a second peak at the time of compression, following reverse-shock interaction. The actual dependence of the escape term as implemented in the model can be translated into an overall dependence as $B_{\rm PWN}^{-1/3} R_{\rm PWN}^{-4/3}$. Because of this scaling and the actual evolution of $B_{\rm PWN}$ and $R_{\rm PWN}$, conditions are less and less favourable to particle escape as time goes by, at least until compression. The first peak in the escape luminosity results from a trade-off between the increasing escape time scale (mostly because the nebula grows) and the rising lepton energy content in the nebula (because of continued particle injection at a high rate while radiative losses diminish). A second peak can occur in cases with sufficient turbulence and results from the strong decrease of the \gls{pwn} size and the simultaneous energization of the particle population upon compression. In all cases, particle escape is so efficient that the nebula is emptied of most of its leptons upon compression. Interestingly, this prevents the appearance of a superefficiency period, during which radiative losses exceed the instantaneous pulsar spin-down. The difference in particle escape luminosities between our low and high turbulence scenarios reaches one to two orders of magnitude. As illustrated in Fig. \ref{fig:res:pwnnrj}, this leads to very different lepton energy content when the nebula gets close to its maximum extent before compression, which will translate into distinctive radiation signatures as we will see below. 

Conversely, the impact of the turbulence prescription on the extent of the \gls{pwn} is very modest as shown in Fig. \ref{fig:res:pwnradius}. Low turbulence setups, and the higher escape they allow, result in a slower growth of the nebula and a later interaction with the reverse shock. The main reason for such a moderate impact can be understood from Figs. \ref{fig:res:pwnlum} and \ref{fig:res:pwnnrj}. In the low turbulence case, particle escape dominates over radiation losses after a few decades to a few centuries, and the particle energy content of the nebula remains below the ordered+turbulent magnetic energy at all times. The latter component therefore determines the pressure in the nebula that drives the expansion of the bounding shell. In the high turbulence case, particle escape becomes a significant loss term only after a few centuries to a few millenia. As a consequence, the particle energy content is much higher, by a factor of a few up to a few tens. Yet, the corresponding pressure only rivals that of the ordered+turbulent magnetic component at late times, when the pulsar has released most of its rotational power, and it never exceeds it by much such that the impact on the \gls{pwn} dynamics remains limited.

We can discuss the impact of neglecting any kick velocity of the pulsar. In the reference model runs, the \gls{pwn} has reached a radius of $\sim$10\pc in $\sim8000$\yr at the onset of compression, with a radius growing faster than linearly in time over that period \citep[$R_{\rm PWN} \propto t^{1.2}$; see][]{VanDerSwaluw:2001}. Consequently, all pulsars with a kick velocity below about 1200\kms would still be contained in the nebula by then, which accounts for the vast majority of the population \citep{Verbunt:2017}. So the kick velocity would not have an influence. Model setups exist for which the pulsar leaves the nebula during compression, implying that particles would be injected almost directly in the remnant after some time, thereby enhancing even more the relative contribution of the \gls{snr} compared to the \gls{pwn} (see Sect. \ref{res_gamlum}).

Figure \ref{fig:res:diffcoef} displays the time evolution of the spatial diffusion coefficient for the different model setups, at a particle energy of 100\tev. For reference, the effective value of the diffusion coefficient at 100\tev for large-scale transport in the Galactic disk is on the order of $10^{30}$\dunit. Turbulence in the nebula as prescribed here results in a diffusion coefficient that is initially four to eight orders of magnitude smaller than in the \gls{ism}. With the decrease of the magnetic field strength and increase of the largest scale of the turbulence, at least up to reverse-shock interaction, the diffusion coefficient progressively rises to about the interstellar value at best, for the model setup with the strongest escape, or to a value that is three orders of magnitude smaller, for the model setup confining particles the most efficiently. 

\begin{figure}[!t]
\begin{center}
\includegraphics[width=0.9\columnwidth]{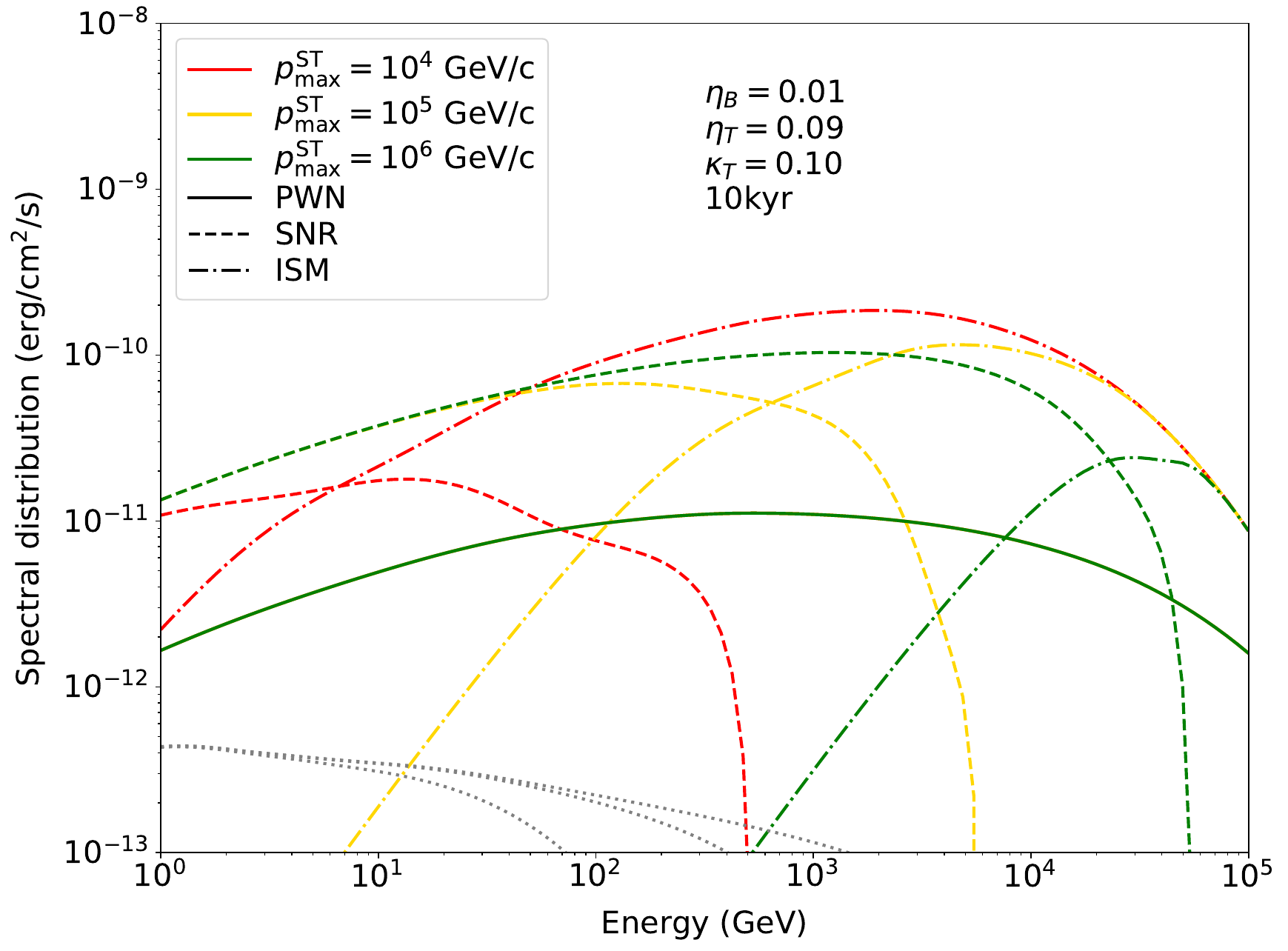}
\caption{Gamma-ray spectra of the \gls{pwn}, \gls{snr}, and \gls{ism} components at 10\kyr in a high-turbulence case, as a function of the maximum particle momentum $p_{\rm max}^{\rm ST}$ allowed in the remnant. The \gls{pwn} emission is the same in all three runs since it is not affected by this parameter. The gray dotted line corresponds to an order-of-magnitude estimate of the pion-decay emission from cosmic rays in the remnant (see text for details).}
\label{fig:res:gamspecpmax}
\end{center}
\end{figure}

\subsection{Gamma-ray luminosity}
\label{res_gamlum}

Figure \ref{fig:res:gamlum} shows the time evolution of the GeV and TeV luminosities of all emission components of the system for four different model setups. In the low turbulence cases, our model predicts that GeV-TeV radiation from the \gls{snr} significantly exceeds that from the \gls{pwn}, by at least about one order of magnitude and up to nearly three orders of magnitude in scenarios with significant escape. For old enough systems, integrated emission from the surrounding \gls{ism} is at least comparable to or much higher than that of the \gls{snr} in the TeV range, while it is largely subdominant in the GeV range past a few decades. The latter statement is however dependent on the parameters $p_{\rm max}^{\rm ST}$ and $\xi$ controlling the time evolution of particle confinement within the remnant. In the high turbulence cases, GeV-TeV emission from the \gls{snr} is unsurprisingly weaker than in the low turbulence case but still predicted to be a significant contribution to the overall signal, with a total flux within a factor of a few of the emission from the \gls{pwn} over most of the valid time range. TeV radiation from the \gls{ism} is also much reduced and becomes comparable to that of the \gls{snr} past a few centuries, while the corresponding GeV radiation remains subdominant. The behaviour observed during compression of the nebula also differs between the low and high turbulence cases: in the low turbulence cases, compression results in an abrupt drop of the \gls{pwn} emission, both in the GeV and TeV ranges, while the high turbulence cases exhibit a modest brightening of the nebula. 

The bulk of the detectable population can be expected to be made up of objects maximizing the age and luminosity product (because the former increases the size of the population, while the latter increases the distance up to which it can be detected). For the reference setups studied here, this would be objects with ages $\sim5-10$\kyr, for which our model predicts a non-negligible if not dominant contribution from the \gls{snr}. In the TeV range, an additional contribution from the surrounding medium can also exist, with a total flux possibly exceeding that of the other components (depending on the parameters of the system), but likely spread over a much larger volume such that its detection may be challenging. Emission from the \gls{pwn} is the dominant contribution only for younger and less evolved objects, with ages ranging from a few centuries to a few millenia, under the assumption of high turbulence (see the red and green curves in Fig. \ref{fig:res:gamlum}). Yet, even in this case, the contribution from the \gls{snr} is at a comparable level at both GeV and TeV energies. This seems consistent with the results obtained for source MSH 15-52 that, at an age of order of 1000\yr, already shows signs of particle escape from the nebula to the remnant \citep{Tsirou:2017}. It may also be consistent with the latest GeV and TeV analysis of the Crab nebula, which reveals a small extension of the source, hardly above the angular resolution limit of the instruments \citep{Aharonian:2024a}: the gamma-ray emission is more extended than the X-ray synchrotron emission, and possibly more than the radio and optical emission especially at GeV energies, which could be indicative of escape out of the nebula.

The bottom panel of Fig. \ref{fig:res:gamlum} shows the predicted luminosities in the $25-100$\tev range. In most model setups considered here, emission in that band is dominated by the \gls{ism} contribution, which may spread over spatial scales much larger than the \gls{snr}+\gls{pwn} system size. Observatories like Tibet AS$\gamma$, HAWC, and LHAASO may therefore be appropriate to search for this emission component and thereby probe particle escape and transport in the vicinity of the source. Interestingly, a large fraction of the sources in the first LHAASO catalogue seem to be positionally coincident with pulsars and very extended, with a 39\% containment radius above 0.5\deg and up to 2\deg, even for the KM2A subsystem \citep{Cao:2024}. The exact \gls{ism} flux level relative to the other components however depends on the details of particle confinement within the remnant. The \gls{ism} luminosity has a characteristic shape resulting from the assumed time evolution of the maximum energy of particles contained in the \gls{snr}. First, the flux in a given band increases until $p_{\rm max}(t)$ rises to a value high enough to confine the emitting particles within the remnant. The supply of fresh particles to the \gls{ism} is then suspended and the emission exhibits a plateau. At a later time, past the Sedov-Taylor time, the forward shock progressively weakens and $p_{\rm max}(t)$ eventually drops down below the value for particles confinement within the \gls{snr} and the \gls{ism} is fed again with new particles, which produces a rise in the emission while at the same time the emission from the remnant decreases as it gets emptied. The duration of the plateau is set by parameters $p_{\rm max}^{\rm ST}$ and $\xi$ and increases with decreasing energy, as can be seen in Fig. \ref{fig:res:gamlum}, where the GeV plateau is much longer than the TeV one.

\subsection{Gamma-ray spectra}
\label{res_gamspec}

Figure \ref{fig:res:gamspec} shows the gamma-ray spectra of each emission component for four different model setups and at three times: close to the maximum lepton content of the nebula (1\kyr), shortly before reverse-shock interaction and after the first peak of particle escape (5\kyr), and at the beginning of compression (10\kyr). Importantly, these spectra are integrated over quite different spatial regions. From 1 to 10\kyr, the \gls{pwn} component arises from a volume increasing from 1 to 10\pc, while the \gls{snr} component is produced in a sphere growing from 5 to 20\pc. The spatial extension of the \gls{ism} component is unclear and will depend on the specifics of particle transport in the vicinity of the remnant: it could be nearly as small as the \gls{snr} in case of strong confinement, or several hundreds of pc across in case of diffusion with the average large-scale properties for the Galactic disk (typically $\sim10^{30}$\dunit at 10\tev). 

In all panels, we provide for comparison a spectrum for pion-decay emission from cosmic rays in the remnant. The latter is computed assuming that, at all times, a constant fraction $5 \times 10^{-7}$ of the material swept-up by the forward shock has undergone diffusive shock acceleration, which produced a power-law distribution in momentum having a constant index 2.3 and extending up to the maximum momentum $p_{\rm max}(t)$ defined by Eq. \ref{eq:pmax}. The injection fraction of $5 \times 10^{-7}$ results in about 8\% of the ejecta kinetic energy being converted into accelerated particles at an age of 10\kyr. These particles interact with gas in the \gls{snr}, swept-up interstellar material and ejecta, assumed to be uniformly distributed within the volume for simplicity. The resulting gamma-ray emission is scaled up by a nuclear enhancement factor of 1.845 \citep{Mori:2009} to account for nuclei in the cosmic-ray population and ejecta. This calculation provides an order-of-magnitude estimate of the emission from the \gls{snr}, for comparison to that powered by particles escaping the \gls{pwn}. It neglects the loss of particles escaping upstream of the forward shock, higher-than-solar metallicity in the ejecta or \gls{ism}, and any radial structure of thermal and non-thermal gas in the \gls{snr}.

The low turbulence spectra in Fig. \ref{fig:res:gamspec} show the dominance of the \gls{snr} and \gls{ism} components over the emission from the \gls{pwn} over 1-10\kyr. The \gls{snr} component is characterized by hard emission up to $0.1-1$\tev, at which energy even harder emission from the \gls{ism} takes over. This crossover energy depends on the parameters controlling confinement in the remnant, and it decreases with time as more and more particles escape from the system. In the high turbulence spectra of Fig. \ref{fig:res:gamspec}, the \gls{pwn} component weighs more in the total signal, and is probably the dominant one in terms of surface brightness as it arises from the smallest volume. With time, the contribution from the \gls{snr} (at $\lesssim$1\tev) and the \gls{ism} (at $\gtrsim1$\tev) gains in intensity such that at ages $\gtrsim5-10$\kyr, it constitutes a non-negligible, if not utterly dominant, fraction of the total integrated signal.

In all model setups, it seems that inverse-Compton emission powered by particles escaping the \gls{pwn} dominates pion-decay emission from cosmic rays accelerated at the forward shock and advected downstream in the \gls{snr}. The pion-decay spectrum provided for comparison is admittedly a simple estimate, and enhancements by a factor of a few are conceivable assuming a higher acceleration efficiency or hadronic interactions in a denser gas layer within the remnant.

Neglecting the \gls{ism} component, assuming for instance rapid particle diffusion after decoupling from the \gls{snr}, the predicted spectra are characterized by a very distinctive pattern that may allow one to indirectly infer the turbulence level: in the low turbulence case, the emission from the system drops off rather abruptly beyond $\sim1-3$\tev, and the \gls{pwn} takes over beyond $\sim10-30$\tev with an intensity that is much lower, about two orders of magnitude or more; conversely, in the high turbulence case, the total \gls{pwn}+\gls{snr} emission is more smooth over the full spectral range. We will see in Sects. \ref{app_1809} and \ref{app_1825} that the latter case seems to be more representative of real systems.

In Fig. \ref{fig:res:gamspecpmax}, we illustrate the effect on the spectra at 10\kyr of the maximum particle momentum $p_{\rm max}^{\rm ST}$ allowed in the remnant, for three values $10^4$, $10^5$, and $10^6$\gevbyc. With increasing maximum momentum, the cutoff in the \gls{snr} emission is pushed to higher and higher energies, while the \gls{ism} emission is restricted to a smaller and smaller energy range at the high end of the spectrum. One should note that the sum of the two components is not constant because allowing particles to escape more rapidly and easily from the remnant preserves them from adiabatic losses.

Although the main goal of this paper is to investigate the effect of escape in \glspl{pwn} on their gamma-ray emission, the model self-consistently predicts synchrotron emission. The corresponding spectra for both low and high turbulence model setups are presented in Appendix \ref{app:radio}, with emphasis on radio emission where our model is more reliable.

\subsection{Gamma-ray morphology}
\label{res_gammorph}

The results discussed above showed that emission from the \gls{snr} may be a non-negligible, if not dominant, contribution to the emission from the system (leaving aside emission from the surrounding \gls{ism}). The exact share depends on the turbulence parameters, and on the energy at which the system is observed. As such, the gamma-ray emission from the whole system may have a specific energy-dependent morphology indirectly conveying information on turbulence. By construction, because of the one-zone assumption for each component of the system, the morphological information that can be extracted from the model is limited. Nevertheless, it provides interesting trends that are reminiscent of some observed patterns.

\begin{figure}[!t]
\begin{center}
\includegraphics[width=0.9\columnwidth]{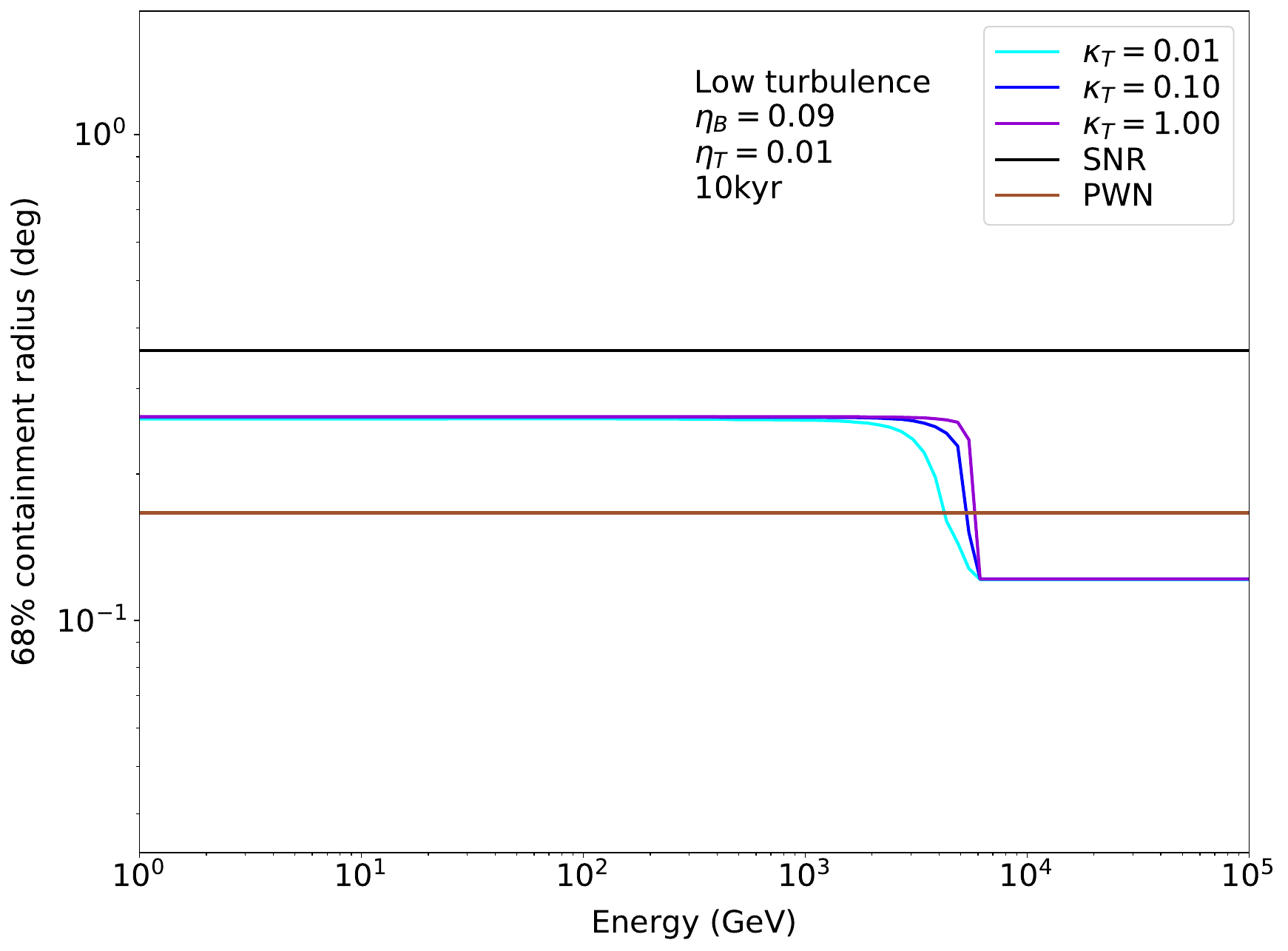}
\includegraphics[width=0.9\columnwidth]{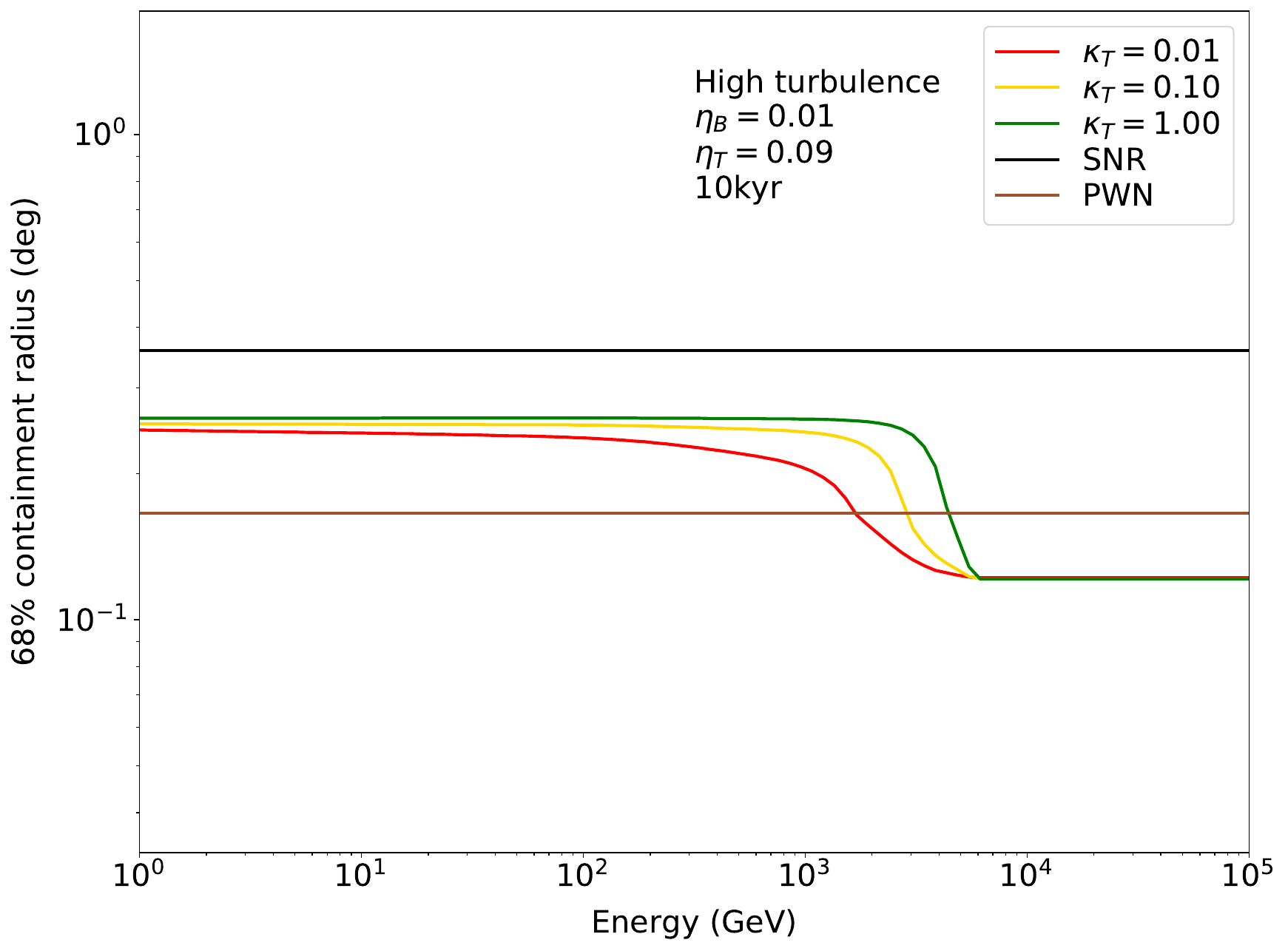}
\caption{The 68\% containment radius of the \gls{pwn}+\gls{snr} emission as a function of energy, for the different model setups and at a system age of 10\kyr. The gray and black dotted lines show the angular extent of the \gls{pwn} and \gls{snr}, respectively, for reference.}
\label{fig:res:gammorph}
\end{center}
\end{figure}

We computed the intensity distribution of the \gls{pwn}+\gls{snr} emission as a function of the angular distance from the centre. Assuming uniform emission properties within each spherical volume, \gls{pwn} or \gls{snr}, we integrated the volumetric emissivity along each line of sight from the inside out (or from the central pulsar to the forward shock), and then assessed the 68\% containment radius of the emission at each gamma-ray energy. The results are displayed in Fig. \ref{fig:res:gammorph}, for the six model setups considered and at a system age of 10\kyr. 

In the low turbulence cases, the \gls{snr} contribution is so strong that the morphology is dominated by the remnant, with a 68\% containment radius at about two thirds of the forward shock extent. This holds until a few TeV, the maximum energy radiated by the highest-energy leptons still contained in the \gls{snr} at this age. Beyond that limit, the \gls{snr} contribution collapses, as illustrated in the bottom panel of Fig. \ref{fig:res:gamspec}, and the morphology becomes dominated by the \gls{pwn}, which leads to a drop by a factor $\sim2$ of the typical size of the emission. In the high turbulence cases, this trend is mitigated by the fact that the \gls{snr} contribution is more comparable to that of the \gls{pwn}. 

Adding the possibility of a contribution from the surrounding \gls{ism}, especially in the case of some confinement in the vicinity the system, would lead to an even richer set of likely morphologies. This is particularly relevant for instruments like HAWC or LHAASO, with their capabilities to detect and image very extended sources above 10\tev.

\section{Application to known sources}
\label{app}

We applied our model on sources HESS J1809$-$193 and HESS J1825$-$137, that we selected for the following reasons: (i) there is in each case an identified pulsar with measured properties (spin period and derivative); (ii) they have extended gamma-ray coverage from GeV to nearly 100\tev; (iii) they are intermediate in terms of dynamical evolution, with ages of order 10-20\kyr, which falls into the applicability domain of our model framework; (iv) they were discussed in past literature as sources for which particle escape is an important aspect, notably because of the large physical size of their gamma-ray emission. 

About point (iii), there are indications from observations that both systems have undergone reverse-shock interactions (strongly asymmetric development of the nebula with respect to the pulsar; see the review of observations in Sects. \ref{app_1809_obs} and \ref{app_1825_obs}). Yet, we caution that it is hard from observations alone to quantify for how long the systems have been in such a stage, and therefore to guarantee that they perfectly fit within the domain of applicability of our model.

About point (iv), this is what justifies the development and use of the three-zone model presented here. One-zone models in their original form, like that introduced in \citetalias{Gelfand:2009}, have difficulties reproducing the physical extent of such sources without stretching the remnant or pulsar parameters to extreme values \citep[e.g. the discussion on HESS J1825$-$137 in][]{DeJager:2009}. Conversely, our model, which allows for the propagation of emitting particles across the whole system out to the surrounding medium, can naturally produce very extended sources (at least qualitatively, since the current version of the model does not compute the radial distribution of emitting particles, especially in the interstellar medium, as emphasised in Sect. \ref{res_caveats}).

\subsection{Fitting strategy}
\label{app_fit}

In our application of the model to HESS J1809$-$193 and HESS J1825$-$137, we will focus on gamma-ray observations only at this stage. The purpose is to obtain a quantitative description of their spectra, and a semi-quantitative description of their morphological properties (because this three-zone model does not yet include a detailed computation of the spatial transport).

The model has a large number of free parameters in total, and some degree of non-linearity, which altogether offers some freedom in fitting data and some dependency on initial values. In order to get meaningful results and a better comprehension of the physical effects, we restricted the fitting procedure to the main subset of the following parameters:
\begin{enumerate}
\item Pulsar spin-down time scale $\tau_0$
\item Ejecta mass $M_{\rm ej}$
\item Ejecta energy $E_{\rm ej}$
\item Interstellar medium density $n_{\rm 0}$
\item Turbulence maximum scale $\kappa_T$
\item Turbulent energy injection efficiency $\eta_{\rm T}$
\item Particle injection spectrum low-energy index $\alpha_1$
\item Particle injection spectrum high-energy index $\alpha_2$
\item Particle injection spectrum break energy $E_b$
\item Particle escape maximum momentum $p_{\rm max}^{\rm ST}$
\end{enumerate}
We emphasise that there is some degeneracy between parameters $\kappa_T$ and $\eta_{\rm T}$, in that the particle escape flux term scales roughly as $\kappa_T^{a}/\eta_{\rm T}^{b}$, where the exponents $a$ and $b$ depend on the turbulence regime. But because there is no strict anti-correlation over the entire parameter space, we kept both quantities in the fitted set.

We aimed at solutions involving pulsar and supernova remnant properties in line with the statistical distributions inferred from population studies. For pulsars, we expected to obtain an initial spin-down power and spin-down time scale yielding an initial spin period and a magnetic field strength in agreement with population studies of radio pulsars \citep{Faucher-Giguere:2006} and gamma-ray pulsars \citep{Watters:2011}. For supernova remnants, we expected $E_{\rm ej}$ and $n_{\rm 0}$ values in agreement with those inferred by \citet{Leahy:2020} from a sample of Galactic X-ray \glspl{snr}, and we aimed at values for $M_{\rm ej}$ in line with those obtained in the explosion models of \citet{Sukhbold:2016}. Similarly, we aimed at getting particle injection properties consistent with the spread of values inferred from spectral studies of a number of established \glspl{pwn} \citep{Torres:2014}.

We emphasise that model complexity and computation time prevent a fine sampling of the large parameter space and a reliable estimation of the uncertainties on the model parameters. For that reason, the best-fit model setups presented below should be understood as sets of values yielding a satisfactory description of the experimental data, rather than measurements of the physical quantities involved. Also, we caution that the domain of applicability of the model (i.e., not too deep into reverse shock crushing of the nebula) tends to favour relatively large ejecta masses, as this delays the \gls{rs}-\gls{pwn} interaction, which in turn implies larger $t_{\rm ST}$ times and smaller $p_{\rm max}^{\rm ST}$ values. In practice, this drove the best-fit $M_{\rm ej}$ values to the maximum acceptable value according to \citet{Sukhbold:2016}, 15\msol. Some of these issues may be solved in future developments of the model.

\subsection{HESS J1809$-$193}
\label{app_1809}

\subsubsection{Summary of observations}
\label{app_1809_obs}

We first applied the model to HESS J1809$-$193, an unassociated gamma-ray source discovered in 2007 and possibly connected to PSR J1809$-$1917, a pulsar with period $P = 0.083$\,s, characteristic age $\tau_{\rm c} = 5.1 \times 10^{4}$\yr and present-day spin-down power of $\dot{E}_{\rm PSR} = 1.8 \times 10^{36}$\punit. The estimated distance to the pulsar is $d=3.3$\kpc \citep{Manchester:2005}. Alternative explanations or complementary contributions such as an \gls{snr} interacting with its molecular environment were put forwards \citep{Castelletti:2016,Araya:2018,Boxi:2024,Li:2023,Aharonian:2023}. A compilation of the multi-wavelength context and detection history can be found in \citet{Aharonian:2023}. 

In the most recent analysis of H.E.S.S. observations by \citet{Aharonian:2023}, HESS J1809$-$193 can be resolved into two morphologically and spectrally distinct components: (i) an extended component A, described spatially by an elliptical 2D Gaussian intensity distribution (with a $1\sigma$ size of 0.6\deg and 0.3\deg along the major and minor axis respectively), and spectrally by a curved spectrum with a cutoff at 13\tev; (ii) a more compact component B, described spatially as a symmetric 2D Gaussian intensity distribution (with a $1\sigma$ size of 0.1\deg) and spectrally by a flat power-law spectrum seemingly extending beyond a few tens of TeV without cutoff. PSR J1809$-$1917 and its X-ray nebula are offset from the centre of the compact component, at a distance of about its characteristic extent, which may result from a combination of asymmetric reverse-shock crushing and pulsar natal kick. HESS J1809$-$193 was detected up to 100\tev and beyond from HAWC observations \citep{Abeysekara:2019}. It is possibly extended at gamma-ray energies $>56$\tev, with a reported $1\sigma$ size of 0.3\deg for a symmetric 2D Gaussian morphology.

In the GeV range, an extended source known as 4FGL 1810.3$-$1925e was reported and is positionally very close to PSR J1809$-$1917 \citep{Abdollahi:2020,Araya:2018}. It has an extension that is intermediate between that of the compact and extended components A and B observed at TeV energies (a $1\sigma$ size of 0.3\deg for a symmetric 2D Gaussian intensity distribution), and a relatively steep spectrum modelled as a simple power law above 1\gev. It seems possible to describe the corresponding GeV emission assuming the morphology of TeV component A, which results in a modest increase in the flux suggesting that most of the emission is indeed captured with the intermediate size morphology \citep{Aharonian:2023}.

\begin{table*}[ht]
\centering
\begin{tabular}{| c | c | c |}
\hline
\celltspace Source & HESS J1809$-$193 & HESS J1825$-$137 \cellbspace \\
\hline
\celltspace Data set & HESS & LAT+HESS \cellbspace \\
\hline
\celltspace Pulsar age $t_{\rm age}$ (yr) & 16475 & 20930 \\
Pulsar spin-down time scale $\tau_0$ (yr) & 34525 & 470 \\
Ejecta energy $E_{\rm ej}$ (\eunit) & $10^{51}$ & $5 \times 10^{50}$ \\
Ejecta mass $M_{\rm ej}$ (\msol) & 15 & 15 \\
Interstellar medium density $n_{\rm 0}$ (\nunit) & 0.1 & 0.01 \\
\textit{Interstellar magnetic field} $B_{\rm ISM}$ ($\mu$G) & 5 & 5 \\
Turbulence scale $\kappa_T$ & 0.02 & 0.01 \\
Turbulent energy injection efficiency $\eta_{\rm T}$ & 0.075 & 0.2 \\
Magnetic energy injection efficiency $\eta_{\rm B}$ & 0.025 & 0.02  \\
Particle energy injection efficiency $\eta_{\rm e}$ & 0.90 & 0.78 \\
Particle injection spectrum low-energy index $\alpha_1$ & 1.2 & 1.0 \\
Particle injection spectrum high-energy index $\alpha_2$ & 2.4 & 2.5 \\
Particle injection spectrum break energy $E_b$ (GeV) & 800 & 210 \\
\textit{Particle injection spectrum cutoff energy} $E_c$ (TeV) & 1000 & 1000 \\
Particle escape maximum momentum $p_{\rm max}^{\rm ST}$ (\gevbyc) & $1.5 \times 10^5$ &  $1.0 \times 10^3$ \\
\textit{Particle escape index} $\xi$ & 2.5 & 2.5 \cellbspace \\
\hline
\celltspace $\chi^2 / N_{\rm dof}$ & 15.2 / 22 & 30.4 / 30 \cellbspace \\
\hline
\end{tabular}
\caption{Summary of the best-fit model setups for HESS J1809$-$193 and HESS J1825$-$137.}
\vspace{0.2cm}
\textit{Notes to the table: Parameters in italics were set and fixed at the beginning of the fitting procedure. In addition, we emphasise that $t_{\rm age}$ was not explicitly fitted and was deduced from $\tau_0$ and $\tau_{\rm c}$. The same applies to $\eta_{\rm e}$, which was deduced from $\eta_{\rm B}$ and $\eta_{\rm T}$. Last, $p_{\rm max}^{\rm ST}$ was fixed to 1\tevbyc for HESS J1825$-$137 because the \gls{snr} was found to have a minor contribution to the signal.}
\label{tab:appli}
\end{table*}

\subsubsection{Best-fit model setup}
\label{app_1809_res}

In the framework of our model, it seems natural to ascribe TeV component B to the \gls{pwn}, while TeV component A could arise either from the \gls{snr} only, or from a combination of the \gls{snr} and the \gls{ism} components. The origin of the GeV emission is less obvious. The source has an extension comparable to that of component A, and a morphology apparently consistent with it, which suggests 4FGL 1810.3$-$1925e could be the GeV counterpart of the total emission from the system (\gls{pwn}+\gls{snr}+\gls{ism}), but the relatively steep spectrum below 10\gev does not seem a priori compatible with that interpretation (see below). Because of this uncertainty on the origin of the GeV emission, we searched for the parameter set best accounting for the TeV spectra only.

The best-fit parameters are given in Table \ref{tab:appli}. The best-fit model involves a system age $t_{\rm age} \simeq 16500$\yr, at which time the \gls{pwn} and \gls{snr} have a radius of 9\pc and 23\pc, respectively. The predicted size for the \gls{pwn} is consistent with the 86\% containment radius of TeV component B (a radius twice the sigma of the fitted 2D Gaussian intensity distribution). The model setup features $M_{\rm ej} = 15$\msol, which leads to a reverse-shock crushing occurring at 15\kyr, such that the \gls{pwn} has been undergoing compression and/or disruption for about two thousand years. This is qualitatively consistent with X-ray observations showing a hard-spectrum nebula extending away from the pulsar in the direction of the TeV source peak \citep{Klingler:2020}.

Enforcing $t_{\rm age} = \tau_{\rm c} - \tau_0$ yields an initial spin-down time scale of 34500\yr. Assuming a constant braking index of 3, applicable for magnetic dipole radiation, this implies an initial spin-down power of $3.9 \times 10^{36}$\punit, an initial spin period of 68\,ms, and an initial magnetic field of $1.7 \times 10^{12}$\,G. These initial properties agree with those inferred from the known pulsar population \citep{Faucher-Giguere:2006,Watters:2011}.

The particle injection spectrum for the \gls{pwn} is a broken power law with index 1.2 below a break energy of 800\gev, and index 2.4 above it, plus a fixed cutoff at 1\pev. A relatively high cutoff is required by the flat spectrum of TeV component A. These values are typical of known young \glspl{pwn} \citep{Bucciantini:2011,Torres:2014}, and so are the efficiencies with 90\% of the pulsar power transferred to relativistic pairs and the remaining 10\% injected predominantly into turbulence. The latter has a maximum spatial scale of 2\% of the \gls{pwn} radius, which is about $1/4$ of the \gls{ts} radius at the current age of the system. 

After diffusive escape from the nebula, following Eq. \ref{eq:pmax}, particles are confined in the remnant up to a maximum momentum of about $p_{\rm max}^{\rm ST}=150$\tevbyc at around $t_{\rm ST} \simeq 4500$\yr and subsequently decreasing in time with a fixed power-law index $\xi = 2.5$. Such a value is consistent with the range inferred for $\gamma-$Cygni from a modeling of the gamma-ray emission in and near the remnant \citep{Acciari:2023}. Yet, the model fitting was performed on TeV data only, so $p_{\rm max}^{\rm ST}$ is not strongly constrained. Lower values by up to one order of magnitude can be considered before the fit to the TeV spectra starts to degrade.

The fitted spectrum in the upper panel of Fig. \ref{fig:res:specfit} shows that HESS J1809$-$193 is dominated by emission from particles spreading out in the \gls{ism} down to an energy of about 1\tev. In that respect, the elongated shape of TeV component A, with a low inclination with respect to the Galactic plane, might indicate anisotropic propagation with faster diffusion along large-scale magnetic field lines that are preferentially aligned with the plane of the Galaxy. The model provides an explanation for the $10-100$\gev emission detected with the Fermi Large Area Telescope (LAT): in that range, the radiation has a flat or hard spectrum and comes primarily from particles that escaped into the \gls{snr} and are still trapped within it. The model cannot account for emission below $10$\gev, which therefore has to be ascribed to another component (for instance the radiation of cosmic rays accelerated at the forward shock, which would account for the steeper spectrum).

An alternative interpretation exists for HESS J1809$-$193, in which the compact component B actually originates from hadronic cosmic rays interacting with nearby molecular clouds \citep{Aharonian:2023}. A solution for such a mixed scenario can easily be found in the present model framework, for instance by using a higher $\kappa_T$ parameter, thereby allowing more particle escape and depressing the spectrum of the \gls{pwn} component, while other parameters are slightly adjusted so that component A still is described as an \gls{snr}+\gls{ism} component.

\subsection{HESS J1825$-$137}
\label{app_1825}

\subsubsection{Summary of observations}
\label{app_1825_obs}

We further applied the model to HESS J1825$-$137. This gamma-ray source is classified as \gls{pwn}, presumably powered by PSR J1826$-$1334 (also known as PSR B1823$-$13), a young pulsar with period $P = 0.1015$\,s, a characteristic age $\tau_{\rm c} = 2.14 \times 10^{4}$\yr and a present-day spin-down power of $\dot{E}_{\rm PSR} = 2.8 \times 10^{36}$\punit. The estimated distance to the pulsar is $d=4$\kpc \citep{Manchester:2005}. A compilation of the multi-wavelength context of the source can be found in \citet{Abdalla:2019}, and we will focus here on its gamma-ray properties. 

In the TeV range, HESS J1825$-$137 is among the brightest sources known, with a flux of about 60\% of the Crab nebula, and one of the most extended, with an emission detected over about 1.5\deg \citep{Abdalla:2019}. The source is highly asymmetric and actually extends predominantly to the southwest of the pulsar. This is interpreted as resulting from an asymmetric reverse-shock interaction, in which the nebula was first hit and compressed on the north-eastern side. This possibility is supported by the discovery of molecular gas north of HESS J1825$-$137 and at a compatible distance \citep{Lemiere:2005}: the evolution of the \gls{snr} in that direction would have occurred more rapidly, with an earlier formation of the reverse shock that pushed parts of the nebula to the southwest and constrained its subsequent evolution to occur in that direction. The TeV emission is markedly energy-dependent, with a spectrum softening with distance from the pulsar, which was interpreted as the progressive cooling of electrons and positrons as they are transported further and further away from the pulsar \citep{Aharonian:2006}. The observed pattern is a rich data set that allows us to constrain the respective contributions of diffusion and advection, and the evolution of the magnetic field within the volume \citep{VanEtten:2011,Collins:2024}. A direct consequence is that source morphology is also strongly energy-dependent, with an emission that shrinks around the pulsar with increasing gamma-ray energy \citep{Abdalla:2019}. The source was also observed with HAWC and detected as extended \citep{Albert:2021}, with a measured spectrum that is above the one inferred from the H.E.S.S. data, by a factor of about two at energies above 5\tev.

In the GeV range, a very significant source is detected and found to be even more extended than in the TeV range \citep{Principe:2020}. The Fermi-LAT observations extend the trend observed with H.E.S.S. and the source size further increases as photon energy decreases, up to nearly 2\deg in radius at 1\gev. The centroid of the emission is also observed to evolve with energy, but the measured pattern is hard to interpret. Spectrally, emission in the Fermi-LAT band is hard and curved. The full SED from 1\gev to 100\tev is well described by a log-parabola or broken power-law shape, with a peak at 100\gev.

\subsubsection{Best-fit model setup}
\label{app_1825_res}

\begin{figure}[!t]
\begin{center}
\includegraphics[width=0.9\columnwidth]{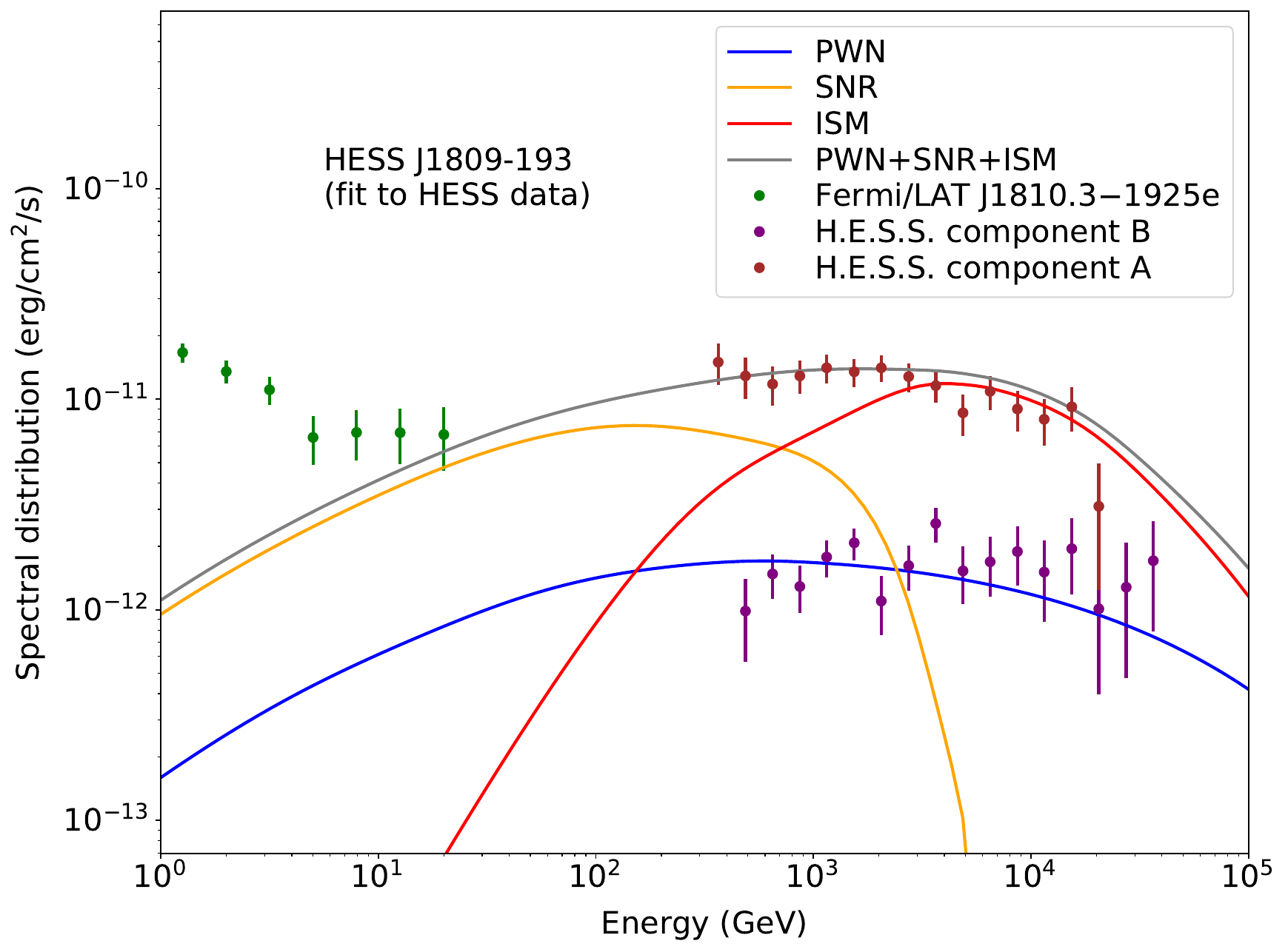}
\includegraphics[width=0.9\columnwidth]{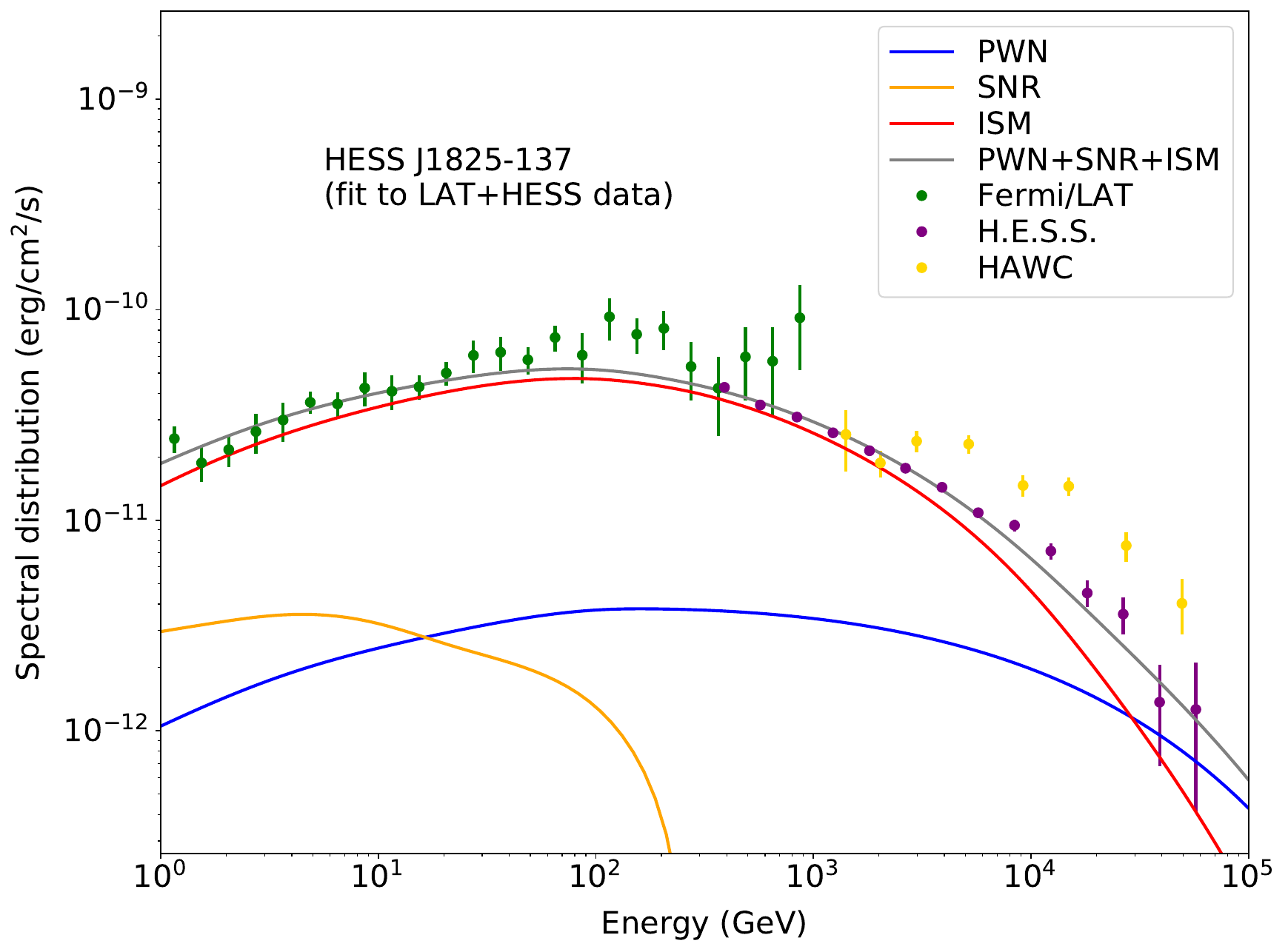}
\caption{Model fits to the spectrum of HESS J1809$-$193 and HESS J1825-137. For HESS J1809$-$193 (top panel), the model was fitted to the H.E.S.S. data points only. The emission below 10\gev is attributed to another component not accounted for in our model framework (see text). For HESS J1825-137 (bottom panel), the model was fitted to Fermi-LAT and H.E.S.S. data points. Upper limits are not shown for clarity but model predictions are consistent with them. In each panel, PWN refers to the emission from particles still trapped in the shocked pulsar wind, SNR refers to the emission from particles that escaped into the remnant and are trapped downstream of the forward shock, and ISM refers to the emission from the halo of particles released in the surrounding medium.}
\label{fig:res:specfit}
\end{center}
\end{figure}

The very large size of HESS J1825$-$137 has been a challenge for most modelling attempts \citep{VanEtten:2011,Khangulyan:2018,Liu:2020,Collins:2024}. Assuming that the observed emission arises from shocked pulsar wind within the nebula implies that the \gls{pwn} has a size of 80\pc at least and the \gls{snr} an even larger size probably exceeding 100\pc (for the southern part of the system). Both requirements can be achieved for a high explosion energy $E_{\rm ej} \gtrsim 3 \times 10^{51}$\eunit, in a very tenuous medium $n_{\rm 0} \lesssim 10^{-3}$\nunit \citep{DeJager:2009,VanEtten:2011}. While such values cannot be excluded, they are hardly compatible with the average properties inferred for Galactic \glspl{snr} \citep{Leahy:2020}. In this work, we take a different approach and aim at a solution based on more typical parameters of the \gls{snr}. As we will see, given the properties of PSR J1826$-$1334, this yields a \gls{pwn} much smaller than 80\pc. Unsurprisingly, the emission observed at large distances from the pulsar is therefore attributed to particles that leaked out of the nebula into the surrounding medium.

Because of the difference between the H.E.S.S. and HAWC spectrum for HESS J1825$-$137, we performed two different model fits: one on Fermi-LAT and H.E.S.S. data, and another one on Fermi-LAT and HAWC data. They yielded very similar parameters sets, the main difference being on pulsar properties. The fit to the Fermi-LAT and H.E.S.S. data is better, in terms of $\chi^2$ given the number of degrees of freedom, and we discuss only this one in the following. The corresponding best-fit parameter set is summarised in Table \ref{tab:appli} and the fitted spectrum is displayed in Fig. \ref{fig:res:specfit}.

The best-fit model setup features a system age $t_{\rm age} \simeq 21000$\yr, at which time the \gls{pwn} and \gls{snr} have a radius of 26\pc and 33\pc, respectively. As illustrated in Fig. \ref{fig:res:specfit}, the gamma-ray emission is dominated by particles spreading out in the \gls{ism} at energies below 30\tev. As discussed further below, the observed extent of the GeV-TeV emission can therefore be used to constrain particle transport in the medium surrounding HESS J1825$-$137. As we approach 100\tev, the contribution from the \gls{pwn} becomes increasingly significant, which is consistent with the H.E.S.S. observation that the emission from a core region of 0.4\deg radius converges towards that of a larger region of 0.8\deg radius above $\sim50-60$\tev \citet{Abdalla:2019}.

The model setup features $M_{\rm ej} = 15$\msol, which leads to a reverse-shock crushing occurring at 12\kyr, such that the \gls{pwn} has been undergoing compression and/or disruption for the past few millenia. We emphasise however that this may hold primarily for the south-western part of the system, while the north-eastern side may have experienced earlier compression due to the presence of molecular material there. We are reaching here the limit of the spherically symmetric model to describe an object that is obviously asymmetric. An initial spin-down time scale of 470\yr is found, which implies an initial spin-down power of $5.8 \times 10^{39}$\punit, an initial spin period of 15\,ms, and an initial magnetic field of $3.3 \times 10^{12}$\,G. These initial properties agree with those inferred from the known pulsar population \citep{Faucher-Giguere:2006,Watters:2011}. The short initial spin period implies that PSR J1826$-$1334 was originally very energetic, with a total rotational energy of $8 \times 10^{49}$\,erg, which may be the reason why HESS J1825$-$137 is such a bright source now. Although this is comparable to the $5 \times 10^{50}$\,erg kinetic energy of the stellar ejecta, it is unlikely that this have a significant impact on the dynamics of the \gls{snr} because more than half of the rotational energy has been radiated away by the nebula after about 1\kyr.

The particle injection spectrum for the \gls{pwn} is a broken power law with index 1.0 below a break energy of 210\gev, and index 2.5 above it, all values that are typical of known young \glspl{pwn} \citep{Bucciantini:2011,Torres:2014}. The injection efficiencies are similar to those inferred for HESS J1809$-$193, with about 80\% of the pulsar power transferred to relativistic pairs and the remaining 20\% going preferentially into magnetic turbulence. The latter is injected with a maximum spatial scale of 1\% of the \gls{pwn} radius, which is about one fourth the radius of the \gls{ts} at the current age of the system. 

The fit to the spectrum, and the large extent of HESS J1825$-$137 over a large GeV-TeV range, imply a largely subdominant contribution from particles trapped in the remnant. In practice, we had to freeze the $p_{\rm max}^{\rm ST}$ parameter, the maximum particle energy reached at $t_{\rm ST} \simeq 14$\kyr, to a low value of 1\tev because it could not be well constrained in the fit. With the fixed parameter $\xi = 2.5$, this means that particles with energies above $\simeq 360$\gev and above should have decoupled from the remnant at the current age of the system, otherwise emission below 10\gev would be dominated by the \gls{snr} component. Overall, in the specific framework of the model, a low $p_{\rm max}^{\rm ST}$ makes it possible for particles to quickly escape into the \gls{ism}, which could be the reason for the puzzling very large extent of HESS J1825$-$137 over such a broad energy range. A more generic interpretation of this requirement is that the parent remnant of HESS J1825$-$137 was relatively inefficient at confining particles for some reason: partial disruption of the \gls{fs} upon interaction with the surrounding material, uneven magnetic conditions over the \gls{fs} surface, maybe in relation to the ambient large-scale magnetic field structure. 

Since our model is not spatially resolved and consists in three concentric and uniform zones, it cannot be quantitatively tested against the measured evolution with distance of the $0.3-5$\tev photon index reported in \citet{Abdalla:2019}. At small angular distances from the pulsar $<0.3-0.4$\deg, the $0.3-5$\tev surface brightness is most likely dominated by the \gls{pwn} contribution that has a relatively flat average spectrum in agreement with observations. As we move to larger angular distances, the emission becomes dominated by the \gls{ism} contribution, and the slightly steeper spectrum predicted by the model is an average of the emission from particles increasingly affected by radiative cooling as they are transported away from the \gls{pwn}, which eventually may be consistent with the observed trend of photon index increasing from about 2 to more than 3 between 0.3\deg and 1.5\deg. This needs to be verified from further development of the model.

Computing the spatial distribution from these particles that escaped the \gls{snr}-\gls{pwn} system and are now experiencing transport in the surrounding \gls{ism} is beyond the scope of this study and will be addressed in a subsequent paper. Several works like those of \citet{VanEtten:2011}, \citet{Liu:2020} and \citet{Collins:2024} have shown the potential of HESS J1825-137 for probing the transport conditions in the medium around the source. We make below some qualitative remarks about this, taking into account the GeV results of \citet{Principe:2020}.

Under the assumption of purely diffusive three-dimensional transport, the typical range of propagating particles can be approximated by:
\begin{equation}
\label{eq:rdiff}
r_{\rm diff}(E) =\sqrt{6 D_{\rm ISM}(E) \times \textrm{min}(t_{\rm age},t_{\rm cool}(E))} \, ,
\end{equation}
where $t_{\rm cool}$ is the cooling time for particles in the \gls{ism}. For a diffusion coefficient with an energy dependence of the form $D_{\rm ISM} \propto E^{\delta}$, and considering leptonic energy losses only in the Thomson limit, this yields $r_{\rm diff} \propto E^{\delta/2}$ for $t_{\rm age} < t_{\rm cool}$ (diffusion-limited regime) and $r_{\rm diff} \propto E^{(\delta-1)/2}$ for $t_{\rm age} > t_{\rm cool}$ (loss-limited regime). The characteristic extent of the gamma-ray emission decreases from about 105 to 15\pc as gamma-ray energy increases from 10\gev to 10\tev. Considering the above scalings, such a monotonic decrease with energy can be explained either by loss-limited diffusion across the whole energy range, or by a peculiar energy dependence of the diffusion coefficient (or a combination of both). The former case can be dismissed because 100\tev particles (radiating typically at $1-10$\tev) are indeed in the loss-limited regime, with $t_{\rm cool} \simeq 4 \times 10^3$\yr, but 100\gev particles (radiating typically at $1-10$\gev) are not, with $t_{\rm cool} \simeq 10^6$\yr. The observed extent of the gamma-ray emission therefore suggests a diffusion coefficient somehow decreasing with energy. Interestingly, such a trend is predicted in calculations of the self-excitation of Alfvenic turbulence by cosmic rays escaping an \gls{snr} and streaming in the hot phase of the \gls{ism} \citep[see][especially their Fig. 4 at an age of $2 \times 10^4$\yr]{Nava:2019}. The levels of diffusion suppression do not quite match, though. At 100\gev, observations indicate a characteristic propagation length of 105\pc over the system age of $2 \times 10^4$\yr, which translates into a diffusion coefficient of $2.6 \times 10^{28}$\dunit. This is about the value assumed as large-scale average coefficient for the Galaxy and consistent with the small levels of diffusion suppression found at these energies in \citet{Nava:2019}. At 100\tev, the characteristic propagation length is 15\pc over a cooling time of $4 \times 10^3$\yr, which translates into a diffusion coefficient of $2.8 \times 10^{27}$\dunit. This is about a factor 300 below the large-scale average value for the Galaxy, and an order of magnitude below the level of diffusion suppression obtained in \citet{Nava:2019}. 

\section{Conclusions}
\label{conclu}

We presented a multi-zone model for the gamma-ray emission from \glspl{pwn}, taking into account the escape of particles out of the nebula into the parent remnant and subsequently to the surrounding \gls{ism}. In its current version, the application of the model framework should be restricted to early evolutionary stages, up to modest levels of compression of the \gls{pwn} upon reverse-shock interaction. The extension to later evolutionary stages will be addressed in a subsequent work, involving numerical hydrodynamical simulations to determine the properties of the remnant.

Relativistic electron-positron pairs injected at the pulsar wind termination shock can escape the nebula as a result of advection and diffusion in the volume. The level and maximum spatial scale of magnetic turbulence in the nebula are the parameters controlling the amount of escape. Particles exiting the nebula are subsequently confined within the remnant because particle acceleration and magnetic field amplification at the forward shock transforms it into a magnetic barrier that keeps energetic particles below a certain maximum momentum downstream of the shock. This confinement is described through a simple time-dependent prescription for this maximum momentum, such that the escape of particles from the \gls{snr} into the surrounding \gls{ism} is controlled via the highest momentum reached in particle acceleration at the forward shock and the power-law index of its subsequent decay in time (see Eq. \ref{eq:pmax}).

For a typical reference system and a wide range of turbulence properties in the nebula, particle escape losses exceed radiative losses after a few centuries. The GeV-TeV emission from the remnant is then comparable to that of the nebula in the case of high turbulence, or completely outshines it in the case of low turbulence. In our reference model setup, this emission dominates the pion-decay radiation from cosmic rays accelerated at the forward shock and advected downstream in the \gls{snr}. In the TeV and especially PeV range, the contribution from particles escaped into the surrounding \gls{ism} has a total flux possibly exceeding by far that of the \gls{snr}+\gls{pwn} components. This is relevant for instruments like Tibet AS$\gamma$, HAWC, and LHAASO, whose capabilities to detect very extended sources above 10\tev make them useful probes of particle escape and transport in the vicinity of the source. 

We applied the model to sources HESS J1809$-$193 and HESS J1825$-$137 with the objective of accounting for the spectral and morphological properties of their gamma-ray emission. The solutions found involve supernova remnants and pulsars with parameters that are typical of the currently known populations, and a nebula that has been undergoing reverse-shock interaction over the past millenia. In the case of HESS J1809$-$193, the compact TeV component is ascribed to the \gls{pwn}, while the more extended TeV component is mostly explained from particles escaped into the \gls{ism}. Particles that escaped the \gls{pwn} but are still trapped in the \gls{snr} can account for most of the hard-spectrum signal detected in the $10-100$\gev range. In the case of HESS J1825$-$137, the broadband GeV-TeV emission is accounted for mostly from particles escaped into the \gls{ism}, while hard-spectrum emission from the \gls{pwn} dominates the signal above $30$\tev. In both cases, high turbulence is required to account for compact hard-spectrum $>10$\tev emission. Despite that, the total emission is completely dominated by particles that escaped into the \gls{ism}. In the framework of our model, this implies that the parent remnants were relatively inefficient particle accelerators, reaching cosmic-ray energies of about $100-150$\tev at most, and maybe far less in the case of HESS J1825$-$137. A more general conclusion, independent of the practical implementation of particle transport in the \gls{snr}, is that the parent remnants were inefficient at confining pairs escaped from the \gls{pwn}, for some reason possibly related to the forward shock structure or integrity. Eventually, the prediction of significant particle escape into the \gls{ism} provides a convenient explanation for the relatively large sizes of the gamma-ray emission in both systems (more than 30\pc in HESS J1809$-$193, and 100\pc in HESS J1825$-$137). In the case of HESS J1825$-$137, our best-fit model setup involves an initially very energetic pulsar, which may be the reason why the source is so bright now.

The model has some potential for the description of other sources exhibiting both compact and extended emission components, for instance MSH 15-52 \citep{Tsirou:2017}, HESS J1813$-$178 \citep{Aharonian:2024b}, or HESS J1834$-$087 \citep{Abramowski:2015}. It also provides in a consistent way the temporal and spectral properties of the flux of particles originally energized by the pulsar wind and ultimately released in the \gls{ism}. It can therefore be used to constrain the escape and transport of particles in the vicinity of pulsar-PWN-SNR systems from broadband gamma-ray observations, or in studies of the contribution of pulsar-related systems to the local electron and positron flux.

\begin{acknowledgements}
The authors acknowledge financial support by ANR through the GAMALO project under reference ANR-19-CE31-0014. This work has made use of the SIMBAD database, operated at CDS, Strasbourg, France, and of NASA's Astrophysics Data System Bibliographic Services. The preparation of the figures has made use of the following open-access software tools: Astropy \citep{Astropy:2013}, Matplotlib \citep{Hunter:2007}, NumPy \citep{VanDerWalt:2011}, and SciPy \citep{Virtanen:2020}. Pierrick Martin wishes to thank Sarah Recchia for useful discussions on particle transport, and Francisco Salesa Greus for providing us with the HAWC data points.
\end{acknowledgements}

\bibliographystyle{aa}
\bibliography{Biblio/Halo.bib,Biblio/ISM.bib,Biblio/ISRF.bib,Biblio/Pulsars.bib,Biblio/DataAnalysis.bib,Biblio/Physics.bib,Biblio/Fermi.bib,Biblio/CosmicRayEscape.bib,Biblio/CosmicRayAcceleration.bib,Biblio/CosmicRayTransport.bib,Biblio/CosmicRayMeasurements.bib,Biblio/GalacticDiffuseEmission.bib,Biblio/SNobservations.bib,Biblio/SNRobservations.bib,Biblio/SNRmodels.bib,Biblio/LMC.bib,Biblio/CTA.bib,Biblio/SNmodels.bib}

\appendix

\section{Detailed summary of original model}
\label{app:mod}

We provide here a description of the original model from \citetalias{Gelfand:2009}, including an introduction of all important quantities and formulae, repeated here for convenience, and clarifications for the few minor variations we implemented. The appendix is organized into the main components of the system.

\subsection{Supernova remnant}
\label{app:mod:snr}

The stellar ejecta interact with a cold circumstellar medium here assumed to have uniform mass density $\rho_0$ and negligible pressure ${\rm P}_0$. This interaction first goes through a non-radiative stage, during which energy losses from thermal radiation of the system are negligible, and then moves towards a radiative stage, where these losses become dynamically important. 

The whole evolution in an idealized setup with spherical symmetry has been the focus of numerous studies over the past decades. In the non-radiative stage, the system rapidly takes the following characteristic structure: the expanding ejecta drive a \gls{fs} propagating in the ambient medium, while a \gls{rs} propagates back in the ejecta; neglecting instabilities, the shocked interstellar medium and shocked ejecta, downstream of the forward and reverse shock respectively, are adjacent and separated by a \gls{cd}. The non-radiative stage is extensively described in \citet[][hereafter TM99]{Truelove:1999}, which provides useful analytical formulae for the dynamics of the \gls{fs} and \gls{rs} as a function of the main parameter of the system\footnote{The prescriptions presented in \citetalias{Truelove:1999} have been critically re-examined in \citet{Bandiera:2021} by comparison to a large set of hydrodynamical simulations covering a wide parameter space. In practice, however, we followed the approach of \citetalias{Gelfand:2009} that relies on \citetalias{Truelove:1999}.}: the ejecta mass $M_{\rm ej}$ and kinetic energy $E_{\rm ej}$, an ejecta density structure parameter $n_{\rm ej}$, and the ambient medium density $\rho_0$.

The ejecta are initially in homologous expansion with velocity being proportional to radius, $v_{\rm ej} (r,t) = r/t$ \citepalias[Eq. 19 in][]{Truelove:1999}, and they remain so until they interact with the \gls{rs}. Their density structure $\rho_{\rm ej} (r,t)$ is defined as a uniform core and power-law envelope with power-law index $n_{\rm ej}$ \citepalias[Eqs. 20 and 24 in][]{Truelove:1999}. The position of the core-envelope transition is set so that the total mass and kinetic energy of the ejecta, given the assumed density and velocity profiles, integrate to $M_{\rm ej}$ and $E_{\rm ej}$. The outer envelope of the ejecta is assumed to be quite steep with $n_{\rm ej} > 5$ \citep[see the discussions and references in][]{Truelove:1999,Bandiera:2021}, our adopted value being $n_{\rm ej} = 9$.

The evolution of the FS in the initial so-called ejecta-dominated stage is described by Eqs. 75-76 with the parameters of Table 6 in \citetalias{Truelove:1999}. As the mass swept-up by the FS becomes significant compared to the ejecta mass, and a significant fraction of the ejecta energy has been transferred to the ambient gas, the dynamics of the FS evolves towards a different scaling with time, described as an offset power-law form of the self-similar Sedov-Taylor solution, defined by Eq. 56 in \citetalias{Truelove:1999}. The transition occurs at time $t_{\rm ST}$ given by Eq. 81, while values for different density structure parameters $n_{\rm ej}$ are listed in Table 6 in \citetalias{Truelove:1999}. The dynamics of the RS is initially obtained from that of the FS via the parameter $\ell_{\rm ED}$, following Eq. 77 in \citetalias{Truelove:1999}. Once the RS enters the core part of the ejecta, its position is described by Eq. 83 in \citetalias{Truelove:1999}.

\subsection{Pulsar wind}
\label{app:mod:psr}

Meanwhile, the central pulsar is assumed to spin down as a result of magnetic dipole radiation. The spin-down luminosity $\dot{E}_{\rm PSR}$ evolves in time as:
\begin{equation}
\label{eq:edotpsr}
\dot{E}_{\rm PSR} = \dot{E}_{\rm PSR,0} \left( 1 + \frac{t}{\tau_0} \right)^{-\frac{n+1}{n-1}} \textrm{ with } n=3 \, .
\end{equation}
The initial spin-down luminosity and spin-down time scale are $\dot{E}_0$ and $\tau_0$, respectively, both of which are related to the initial values of the spin period and magnetic field $P_0$ and $B_0$:
\begin{equation}
\label{eq:edotzero}
\dot{E}_{\rm PSR,0} = \frac{2 \pi^2 I_{\rm NS}}{\tau_0 P_0^2 } = \frac{8 \pi^4 B_0^2 R_{\rm NS}^6}{3 c^3 P_0^4} \, .
\end{equation}
$I_{\rm NS}$ and $R_{\rm NS}$ are the moment of inertia and radius of the neutron star, respectively, for which we take typical values of $10^{45}$\,g\,cm$^2$ and 12\,km.

Under the assumption that spin-down results from magnetic dipole radiation and that the perpendicular component of the magnetic dipole moment does not change significantly with time, the spin-down time scale is related to the pulsar true age $t_{\rm age}$ and its characteristic age $\tau_{\rm c}$ as:
\begin{equation}
\label{eq:taupsr}
t_{\rm age} = \tau_{\rm c} - \tau_0 = \frac{P}{2\dot{P}} - \frac{3 c^3 P_0^2 I_{\rm NS}}{4 \pi^2 B_0^2 R_{\rm NS}^6} \, .
\end{equation}
In the limit that $P \gg P_0$, the age of the pulsar can be approximated from the measurable quantity $\tau_{\rm c} = P / 2\dot{P}$.

The spin-down powers a relativistic outflow, the pulsar wind, that, at sufficiently large distance from the pulsar, is mostly composed of a toroidal magnetic field and electron-positron pairs. Upon interaction with the surrounding medium, the wind is halted at a \gls{ts}, where particles are efficiently accelerated up to very high energies. The acceleration mechanism is not yet fully elucidated and may involve several sites and processes \citep[diffusive shock acceleration at the TS, turbulent acceleration downstream of it, etc; see][]{Bucciantini:2011,Amato:2020}, but observations indicate it drains a large fraction of the wind kinetic energy, of the order of several tens of percent \citep{Zhang:2008,Bucciantini:2011,Torres:2014}.

We neglect the possibly complex radial and latitudinal dependence of the pulsar wind and assume an isotropic outflow. The \gls{ts} forms where the wind ram pressure equals the pressure of the surrounding medium \citep{Gaensler:2006}:
\begin{equation}
\label{eq:rts}
R_{\rm TS} (t) = \sqrt{\frac{\dot{E}_{\rm PSR} (t)}{4 \pi c P_{\rm PWN} (t)}} \, .
\end{equation}
The surrounding medium is initially the \gls{snr}, but very rapidly becomes the \gls{pwn} itself, that is downstream of the shock.

Energy is injected in the nebula in the form of non-thermal pairs and magnetic field. In the original model:
\begin{align}
\dot{E}_{\rm inj} (t) &= \dot{E}_{\rm inj,B} (t) + \dot{E}_{\rm inj,e} (t) = \dot{E}_{\rm PSR} (t) \, , \\
&= \eta_{\rm B} (t) \dot{E}_{\rm PSR} (t) + \eta_{\rm e} (t) \dot{E}_{\rm PSR} (t) \, .
\end{align}
We neglect a possible ionic component and assume constant injection efficiencies.

Pairs are assumed to have a spectrum at injection $Q_{\rm inj,e}$, as a function of particle kinetic energy $E$, consisting of a broken power-law distribution with exponential cutoff :
\begin{equation}
\label{eq:injspec}
Q_{\rm inj,e} (E,t) = Q_0(t) \times e^{-E/E_{\rm c}} \times \begin{cases} (E/E_{\rm b})^{-\alpha_1} & \mbox{if } E\leq E_{\rm b} \, , \\
(E/E_{\rm b})^{-\alpha_2} & \mbox{if } E > E_{\rm b} \, .
\end{cases}
\end{equation}
The modelling of the broadband emission from observed PWNe suggest $\alpha_1$ values in the range $1.0-2.0$, $\alpha_2$ values in the range $2.0-2.8$, and $E_{\rm b}$ in the range $100-500$\gev \citep{Bucciantini:2011,Torres:2014}. We assume a constant cutoff energy $E_{\rm c}$ with typical values in the range $0.1-1$\pev. For relatively old systems, with ages of a few 100\,kyr, the time evolution of the potential drop can be expected to shift the maximum attainable energy to below 100\tev \citep{DeOnaWilhelmi:2022}.

In practice, we consider a kinetic energy grid running from 0.1\gev to 1\pev, and the injection spectrum integrated over that range yields:
\begin{equation}
\label{eq:injpow}
\dot{E}_{\rm inj,e} (t) = \int_{{\rm{0.1\gev}}}^{{\rm{1\pev}}} E Q_{\rm inj,e}(E,t) dE \, .
\end{equation}
Since the pair injection efficiency is assumed constant, the pair injection spectrum is normalized at each time by the pulsar spin-down power.

\subsection{Pulsar wind nebula}
\label{app:mod:pwn}

The \gls{pwn} is the volume $V_{\rm PWN}$ excavated by the shocked pulsar wind as it expands. It is bounded on the inner side at a radius $R_{\rm TS}$ by the \gls{ts}, and on the outer side at a radius $R_{\rm PWN}$ by a thin shell of swept-up ejecta material. The model treats the nebula as a single zone, in which quantities are assumed to be uniform.

The dynamics of this shell is controlled by its inertia and the pressure difference between the nebula on the inner side, and the stellar ejecta on the outer side. At early times, the innermost unshocked stellar ejecta are assumed to have a negligible pressure because of their fast expansion and rapid cooling. Later on, the reverse shock propagating back through the ejecta will eventually reach the nebula. The material ahead of the bounding shell is heated and compressed and the pressure immediately outside the nebula becomes non-negligible, and possibly well in excess of that in the nebula, thereby leading to a strong compression of the nebula. We assume that past this reverse-shock crushing occurring at time $t_{\rm crush}$, the remnant rapidly settles into a state described by the Sedov-Taylor solution \citep[but see][]{Bandiera:2023b}. The conditions ahead of the shell are therefore:
\begin{equation}
\label{eq:pej}
P_{\rm ej} (R_{\rm PWN},t) = \begin{cases} 0.0 & \mbox{if } t \leq t_{\rm crush} \, , \\
\bar{P}_{\rm ST} (R_{\rm PWN}/R_{\rm FS}) P_{\rm FS}(t) & \mbox{if } t > t_{\rm crush} \, .
\end{cases}
\end{equation}
\begin{equation}
\label{eq:rhoej}
\rho_{\rm ej} (R_{\rm PWN},t) = \begin{cases} \rho_{\rm ej} (R_{\rm PWN},t) & \mbox{if } t \leq t_{\rm crush} \, , \\
\bar{\rho}_{\rm ST} (R_{\rm PWN}/R_{\rm FS}) \, . \, \rho_{\rm FS}(t) & \mbox{if } t > t_{\rm crush} \, .
\end{cases}
\end{equation}
Quantities $P_{\rm FS}(t)$ and $\rho_{\rm FS}(t)$ are the pressure and density immediately downstream of the FS and the terms with the ST subscripts are the self-similar Sedov-Taylor profiles, computed following \citet{Bandiera:1984}. In practice, for typical values of the system's parameters such as those used in Sect. \ref{res}, the reverse-shock crushing occurs at $t_{\rm crush} > t_{\rm ST} > t_{\rm core}$, which justifies the above assumption.

The bounding shell mass $M_{\rm PWN} (t)$ is contributed to by ejecta material collected when shell velocity $v_{\rm PWN}(t)$ exceeds the ejecta velocity immediately ahead of the shell, $v_{\rm ej} (R_{\rm PWN},t)$ \citep[we use here a corrected version of the original formula following footnote 1 in][]{Bandiera:2023b}. The shell width is neglected, an assumption justified by one-dimensional hydrodynamical simulations that show a shell width of the order of two percent of the shell radius past a few centuries \citep{Jun:1998}. Two-dimensional calculations show that the shell is susceptible to a series of flow instabilities that lead to some mixing and thickening of the shell \citep{Jun:1998,Porth:2014b}. Past reverse-shock interaction, the thin-shell assumption becomes highly questionable \citep{Bandiera:2023b}. 

The equations governing the evolution of the nebula are detailed in \citetalias{Gelfand:2009}, and we just recall here the main ones. The dynamics of the bounding shell is driven by the pressure difference between the interior of the nebula and the ejecta immediately ahead of the shell, with a possible contribution from the momentum of swept-up mass:
\begin{equation}
\label{eq:mvpwn}
\frac{d}{dt} \left( M_{\rm PWN} v_{\rm PWN} \right) = 4 \pi R_{\rm PWN}^2 \left( P_{\rm PWN} - P_{\rm ej} (R_{\rm PWN}) + C_{\rm PWN} \right) \, .
\end{equation}
The last term on the right-hand side is the contribution from the momentum of swept-up mass and is included only if $v_{\rm PWN}(t) > v_{\rm ej} (R_{\rm PWN},t)$. The pressure inside the nebula $P_{\rm PWN}$ is contributed to by magnetic field and relativistic pairs:
\begin{equation}
\label{eq:ppwn}
P_{\rm PWN} = P_{\rm PWN,B} + P_{\rm PWN,e} + P_{\rm PWN,T} \, .
\end{equation}

The evolution of the magnetic pressure is computed following \citetalias{Gelfand:2009}, while the pressure from relativistic pairs derives from their total energy in the volume, assuming an adiabatic index of 4/3:
\begin{equation}
\label{eq:pe}
P_{\rm PWN,e} (t) = \frac{E_{\rm PWN,e} (t)}{3 V_{\rm PWN} (t)} = \frac{1}{3 V_{\rm PWN} (t)} \int_{{\rm{0.1\gev}}}^{{\rm{1\pev}}} E N_{\rm e}(E,t) dE \, .
\end{equation}
The spectral distribution of pairs at a given time, $N_{\rm e}(E,t)$, is obtained by solving a transport equation involving injection and continuous energy losses:
\begin{equation}
\label{eq:ne}
\frac{\partial N_{\rm e}}{\partial t} + \frac{\partial \dot{E}_{\rm loss,e} N_{\rm e}}{\partial E} = Q_{\rm inj,e} \, .
\end{equation}
The particle energy loss term $\dot{E}_{\rm loss,e}$ includes radiative losses from inverse-Compton scattering and synchrotron radiation as well as losses from adiabatic expansion of the nebula.
The expression for energy losses from radiative processes including inverse-Compton scattering in the Klein-Nishina regime was approximated following \citet{Moderski:2005}:
\begin{align}
\label{eq:tcool}
\dot{E}_{\rm rad,e} &= \dot{E}_{IC} +\dot{E}_{syn} \, , \\
&= -\frac{4}{3}c\sigma_T \gamma^2\left[ \sum_i\frac{U_{i}}{(1+4 \gamma \epsilon_i)^{3/2}} + \frac{B_{\rm PWN}^2}{8\pi}\right] \, ,
\end{align}
where $\sigma_T$ is the Thomson cross-section and $\gamma$ the Lorentz factor of the particle. The sum is performed over all components of the radiation field, described as graybodies with energy densities $U_i$ and temperatures $T_i$, yielding normalized photon energies for the fields $\epsilon_i = 2.8k_BT_i/m_ec^2$, where $k_B$ is the Boltzmann constant, $m_e$ the electron mass, and $c$ the speed of light.

\subsection{Numerical aspects}
\label{app:mod:num}

We implemented the model following Sect. 2.2 in \citetalias{Gelfand:2009}, using a logarithmic grid in both time and particle energy, and starting from the initial conditions exposed in their Appendix B. Given the rather crude numerical scheme, small enough time step is required, of the order of the few thousands steps per decade at least, in order to properly capture the compressions and re-expansions of the nebula after reverse-shock interaction (this has however a limited impact here since our results and discussions focus on the early stage, until modest levels of compression).

\section{Investigation of time scales involved}
\label{app:time}

We assessed some of the hypotheses subtending our model by computing the relevant time scales. As a reference, we use the dynamical time scale for the evolution of the nebula, which is:
\begin{equation}
\label{eq:tdyna}
\tau_{\rm dyna} = \frac{R_{\rm PWN}}{v_{\rm PWN}} \, .
\end{equation}

\begin{figure}[!h]
\begin{center}
\includegraphics[width=0.9\columnwidth]{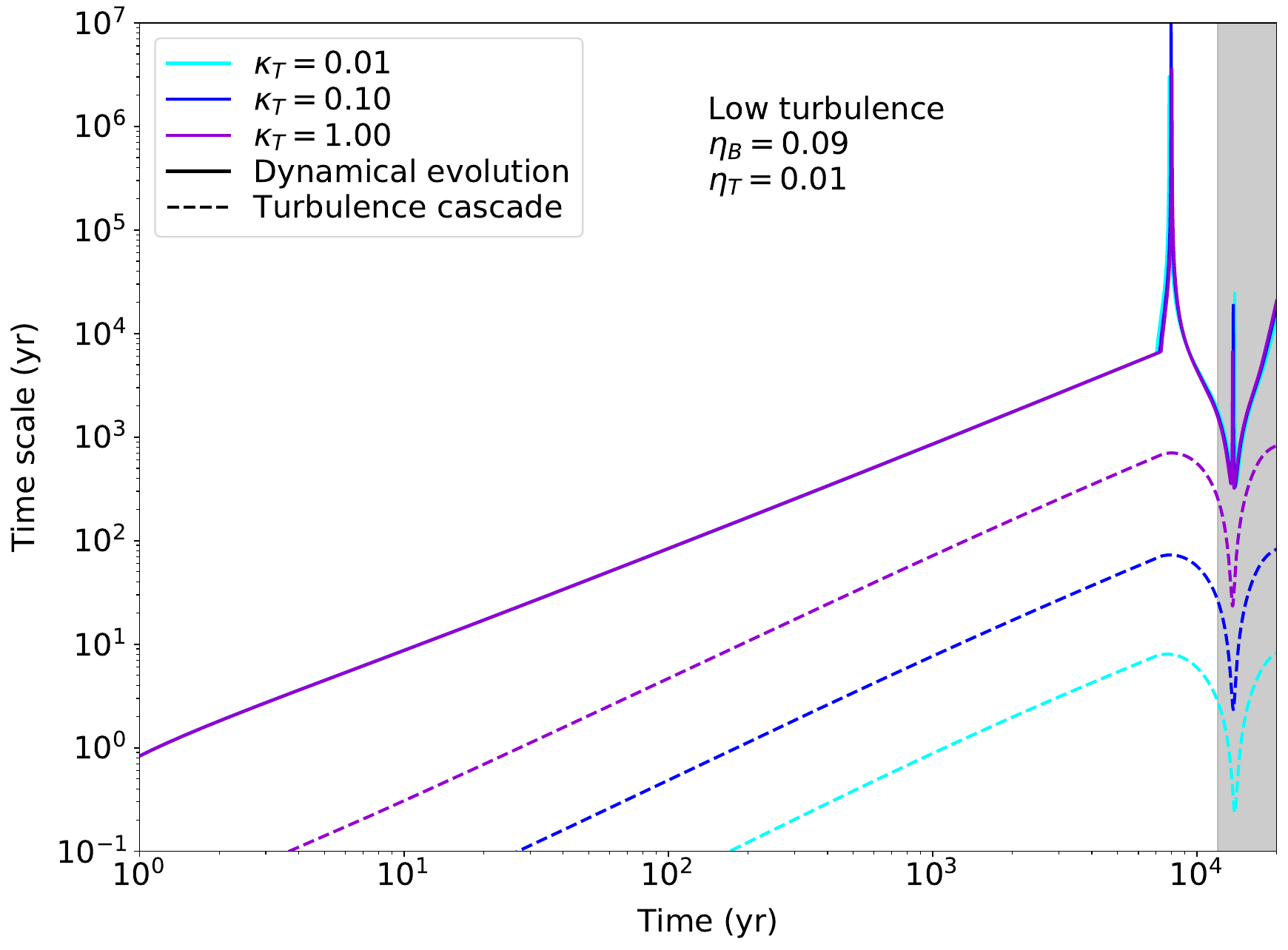}
\includegraphics[width=0.9\columnwidth]{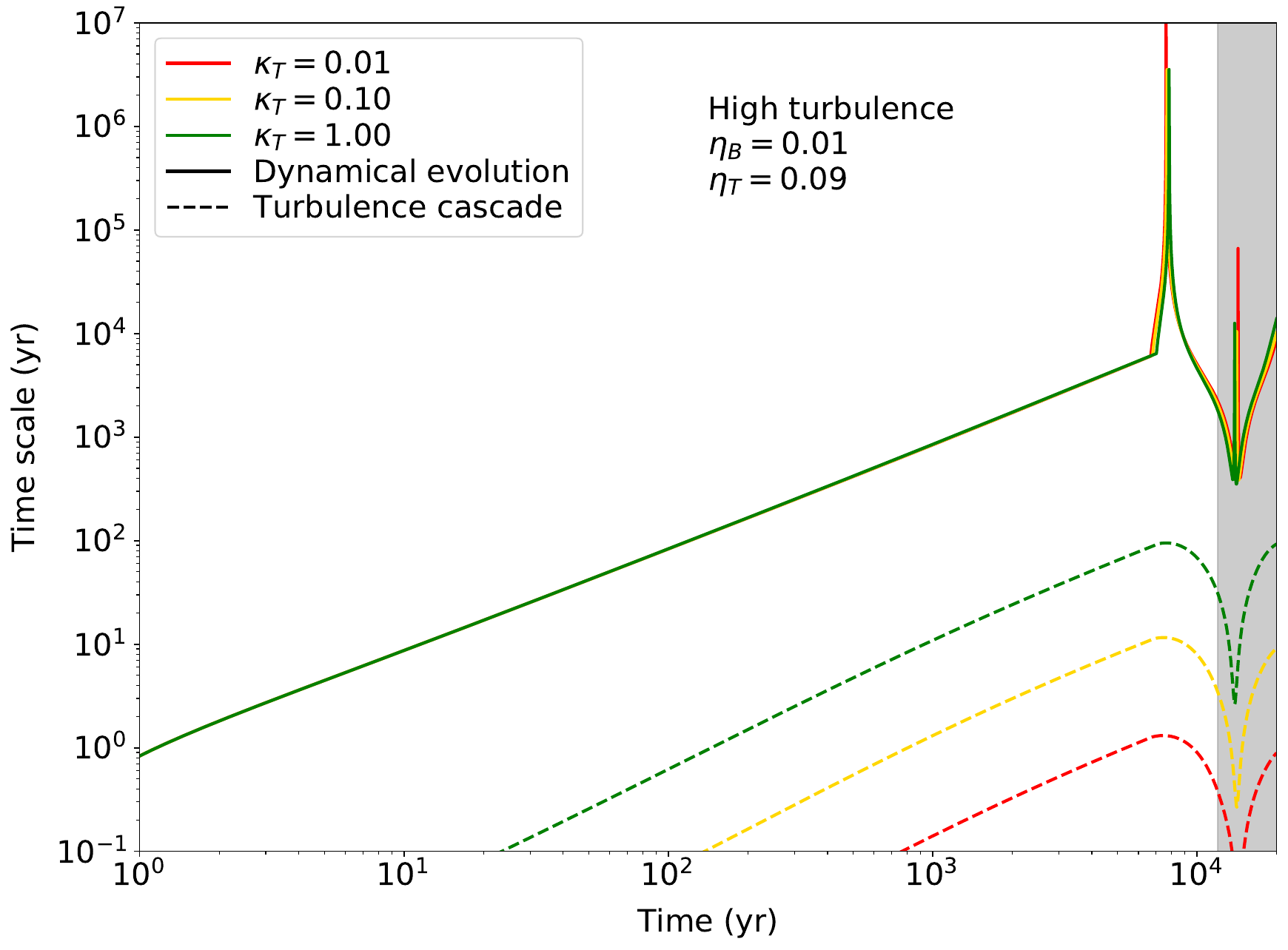}
\caption{Time scales of the dynamical evolution of the nebula and the turbulence cascade in the different model setups.}
\label{fig:app:times}
\end{center}
\end{figure}

We first check the assumption that turbulence is fully developed at all times. We computed the time scale for diffusion in wavenumber space, as a result of non-linear wave-wave interactions, which results in the so-called cascade of turbulence from small wavenumbers (large physical scales) to large wavenumbers (small physical scales). The corresponding coefficient for a Kolmogorov phenomenology is \citep{Miller:1995}:
\begin{equation}
\label{eq:dkk}
D_{\rm kk} = 0.052 v_A k^{7/2} W(k)^{1/2} \, .
\end{equation}
In steady-state, this diffusion coefficient yields a Kolmogorov spectrum, $w(k) \propto k^{-5/3}$, with constant energy flux over the relevant wavenumber range. The corresponding time scale is:
\begin{equation}
\label{eq:tcasc}
\tau_{\rm casc} = \frac{k^2}{D_{\rm kk}} \, .
\end{equation}
This time is a decreasing function of $k$ for a Kolmogorov spectrum, $\tau_{\rm k} \propto k^{-2/3}$, so we evaluated the time scale at the largest spatial scale. The results are plotted in Fig. \ref{fig:app:times}. Unsurprisingly, the cascade proceeds all the more rapidly that the level of turbulence is high and its largest spatial scale is small\footnote{Decreasing the energy-containing scale by some amount $x$ decreases the spectral density at this scale by the same amount $x$ (because turbulent magnetic energy $\sim k_{\rm min} W(k_{\rm min})$); the diffusion coefficient is therefore increased by $x^3$ and the cascading time scale at the energy-containing scale is eventually decreased by $x$.}.

As can be seen from Fig. \ref{fig:app:times}, the turbulence cascade time scale is at least one order of magnitude smaller than the dynamical time scale in our model setups, which justifies our assumption of fully-developed turbulence at all times.

\section{Radio synchrotron emission}
\label{app:radio}

\begin{figure}[!t]
\begin{center}
\includegraphics[width=0.9\columnwidth]{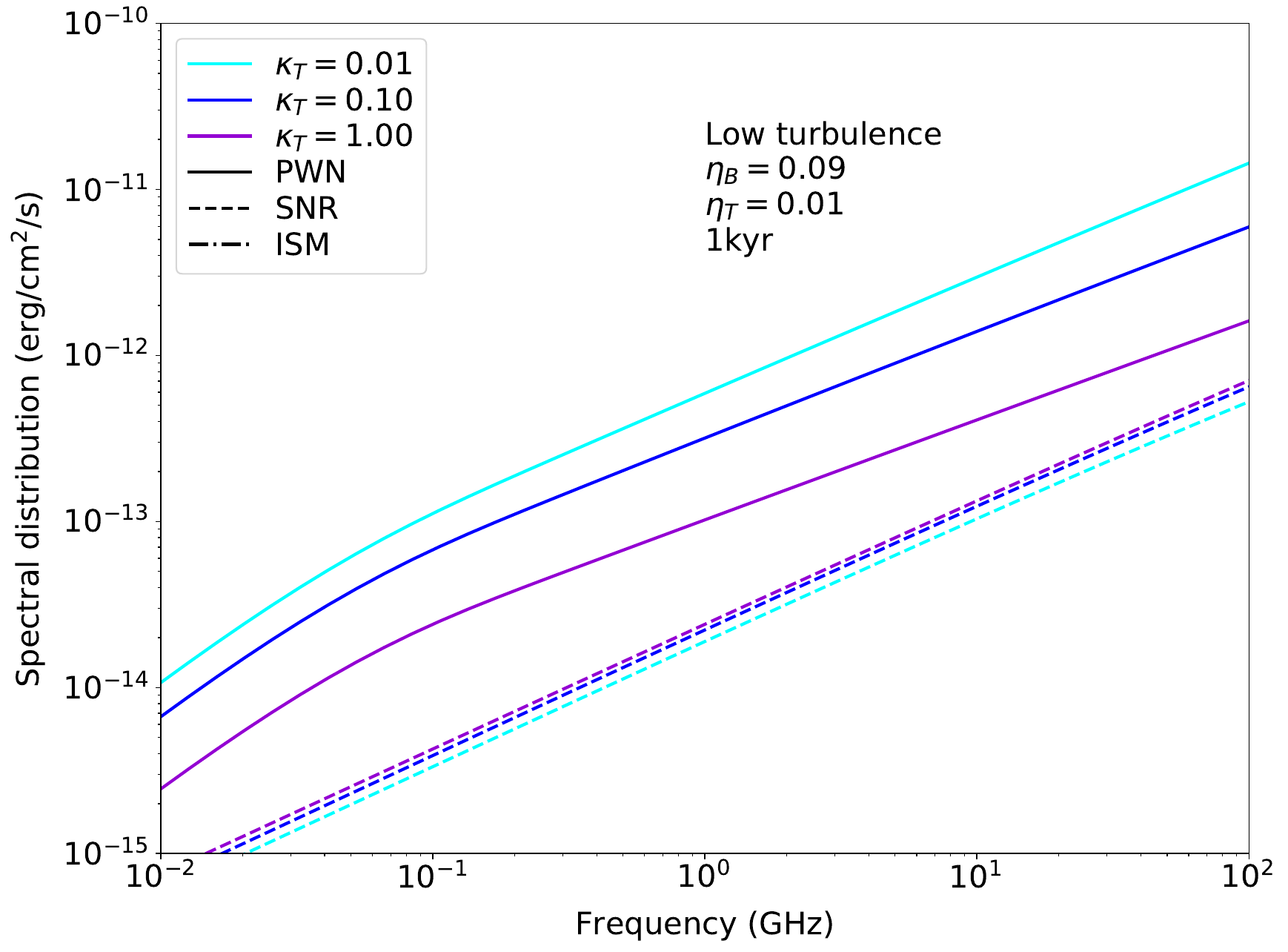}
\includegraphics[width=0.9\columnwidth]{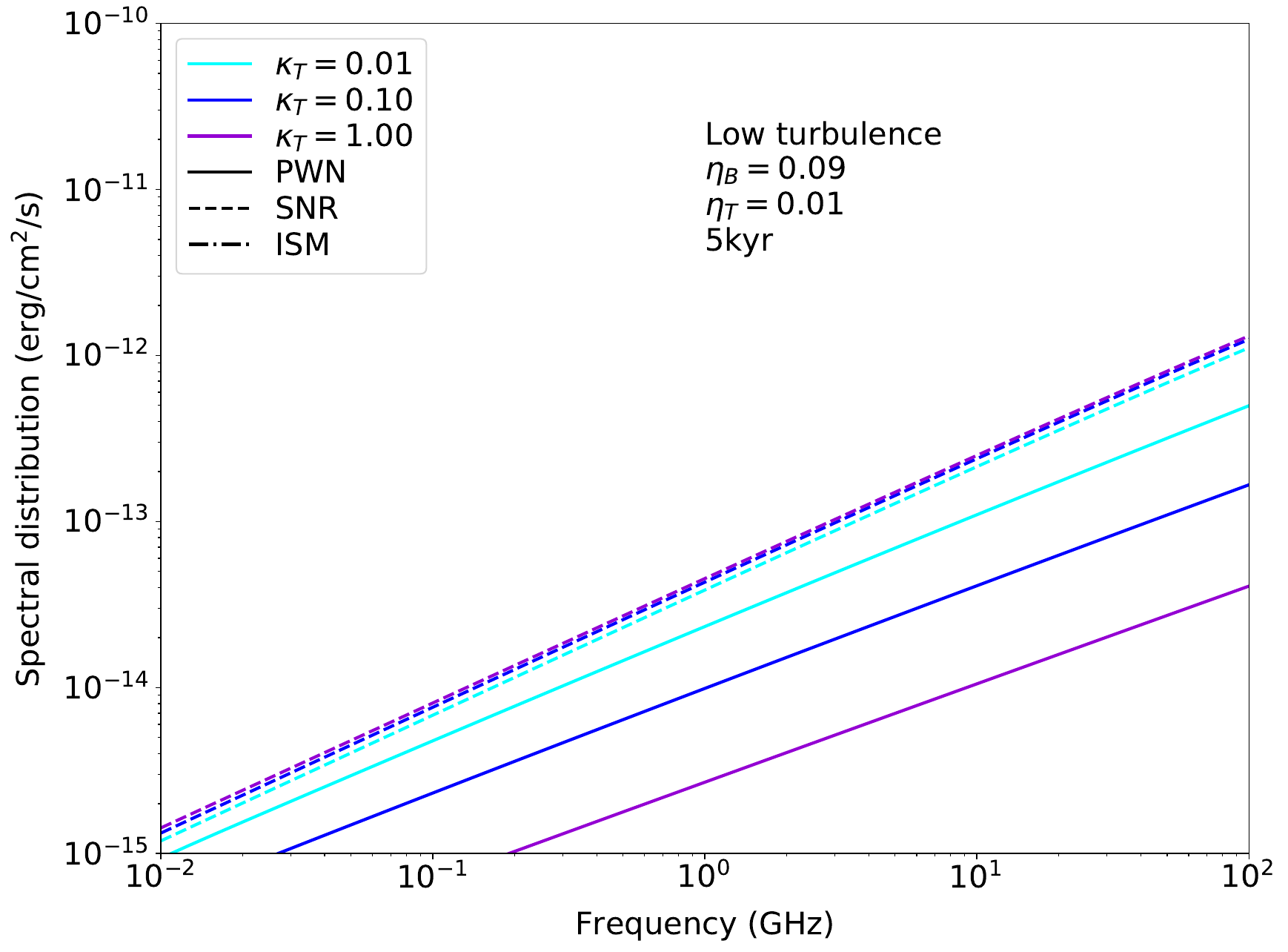}
\includegraphics[width=0.9\columnwidth]{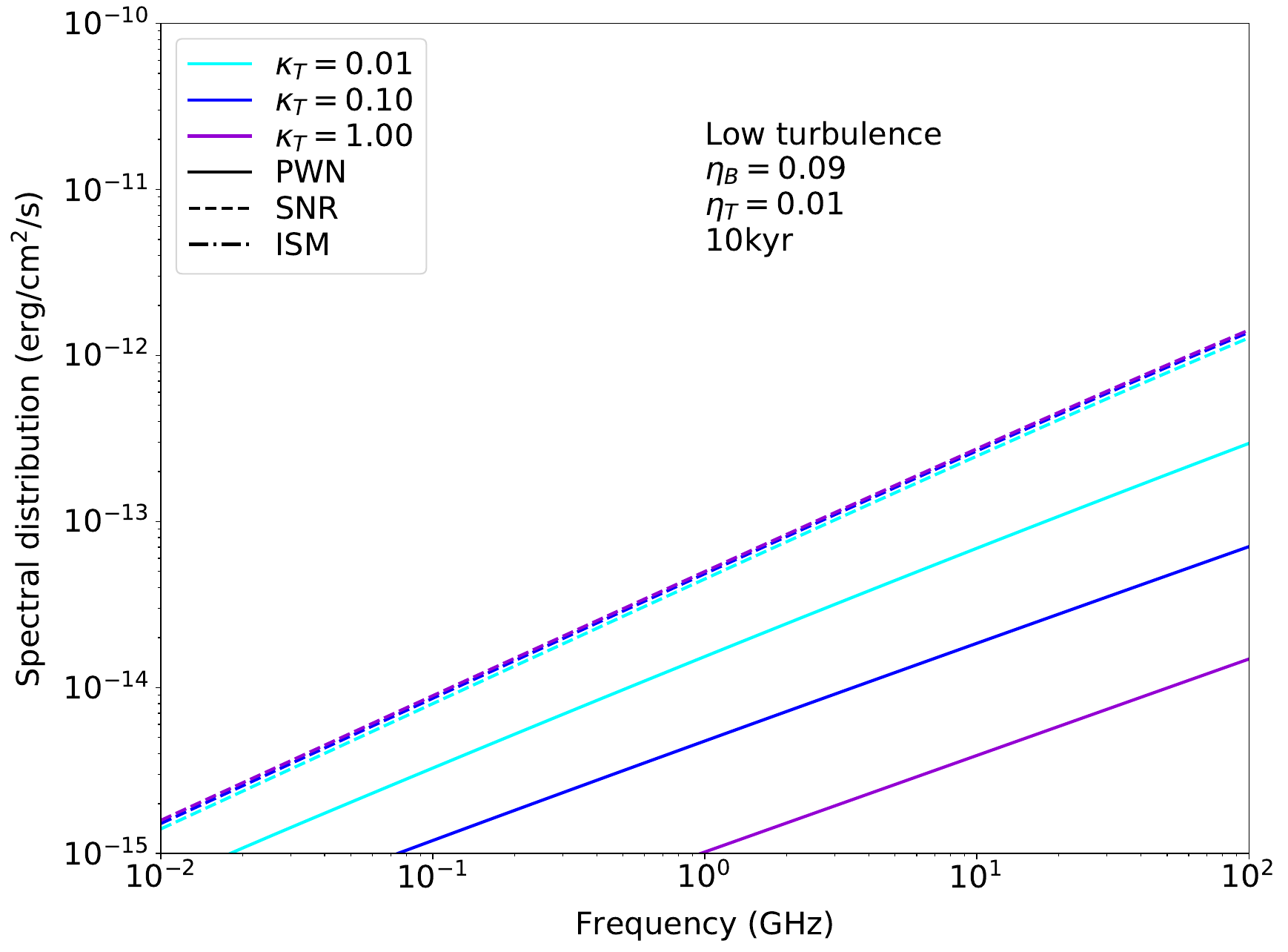}
\caption{Radio synchrotron spectra of the \gls{pwn}, \gls{snr}, and \gls{ism} components in low turbulence model setups at different ages.}
\label{fig:res:radspeclow}
\end{center}
\end{figure}

\begin{figure}[!t]
\begin{center}
\includegraphics[width=0.9\columnwidth]{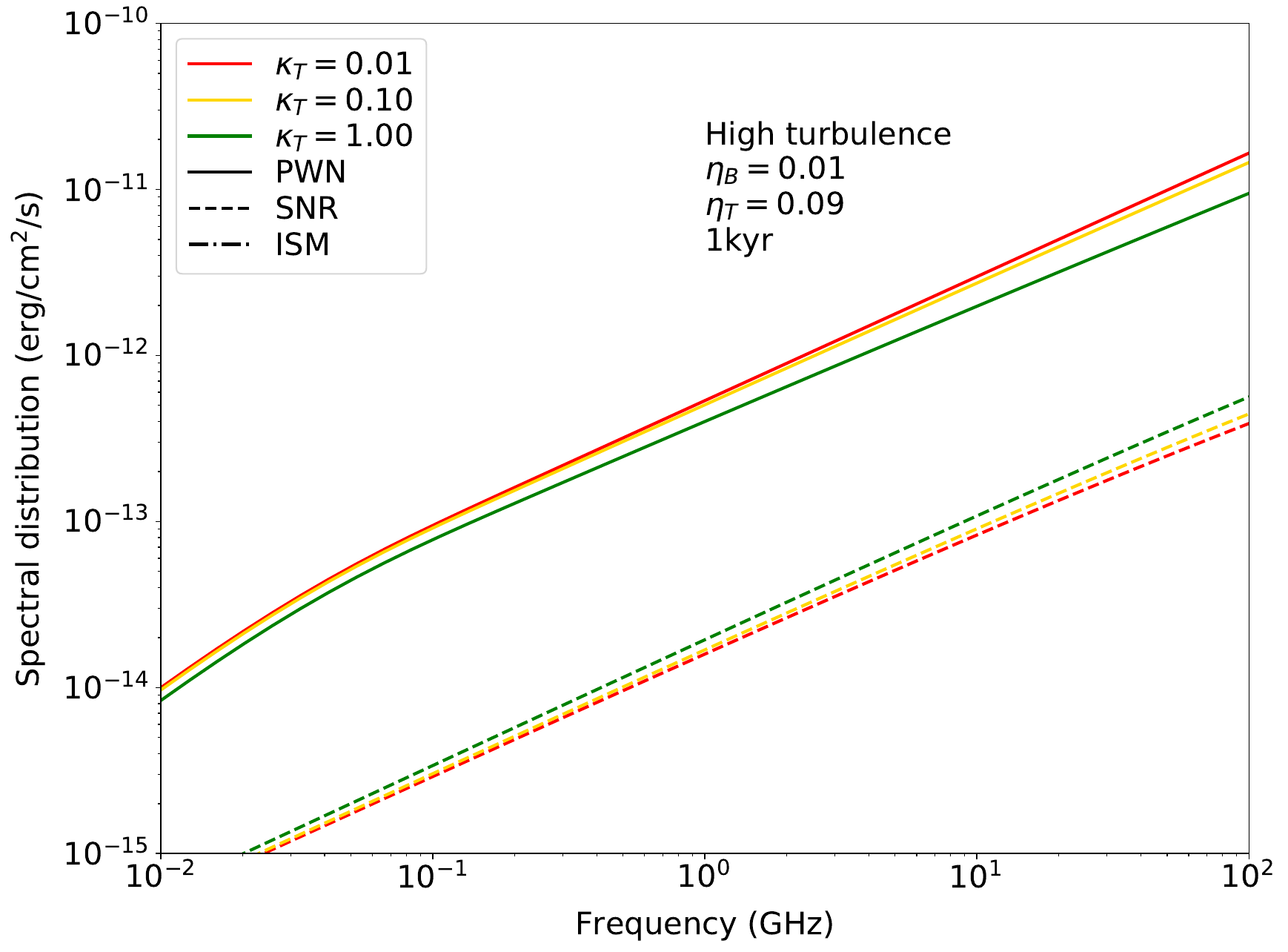}
\includegraphics[width=0.9\columnwidth]{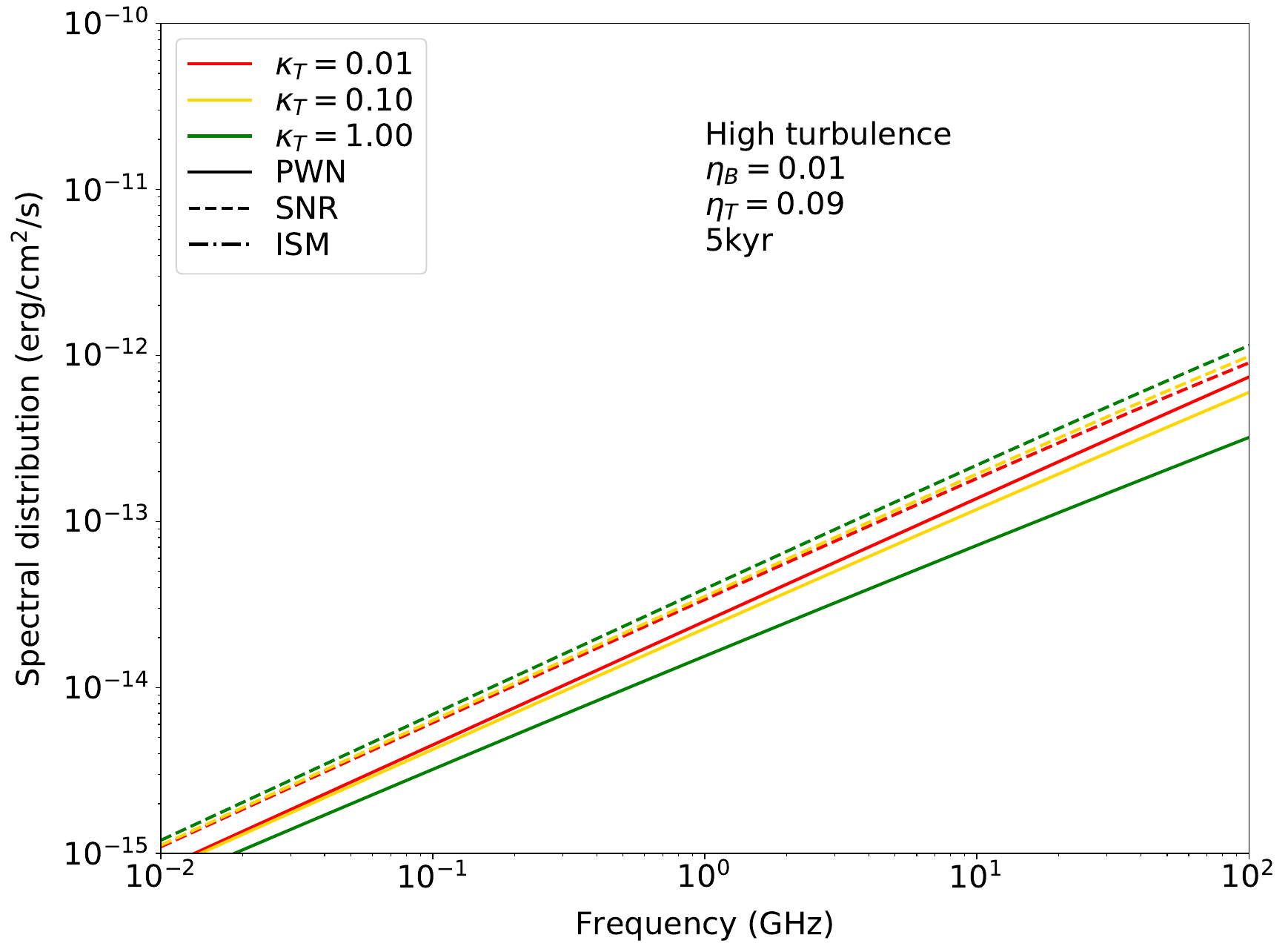}
\includegraphics[width=0.9\columnwidth]{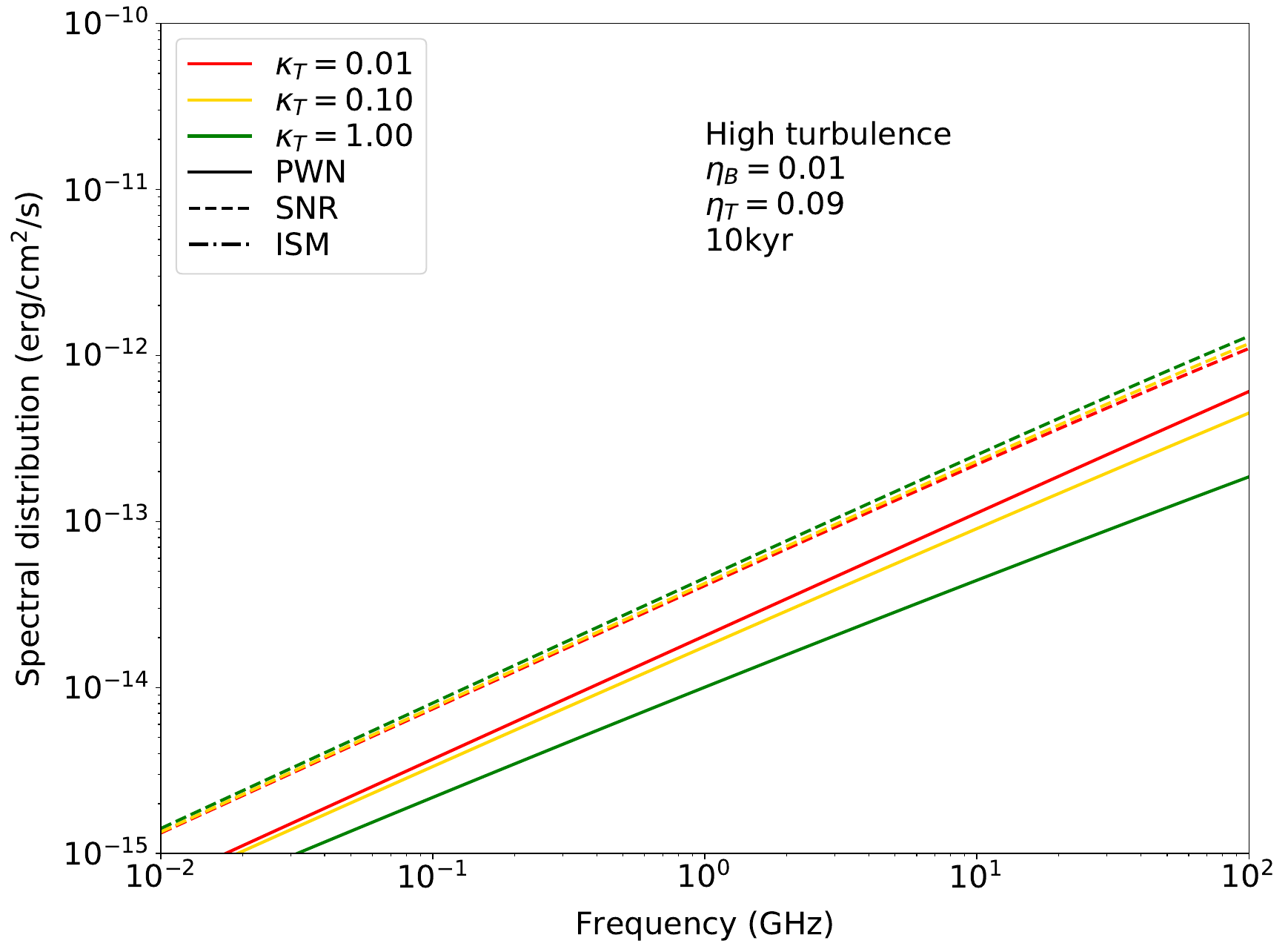}
\caption{Radio synchrotron spectra of the \gls{pwn}, \gls{snr}, and \gls{ism} components in high turbulence model setups at different ages.}
\label{fig:res:radspechigh}
\end{center}
\end{figure}

Figures \ref{fig:res:radspeclow} and \ref{fig:res:radspechigh} show the radio synchrotron spectra of each emission component for the different model setups and at the same three times as before. The magnetic fields in the nebula at these times are in the range $120-180$\mug, $70-110$\mug, and $55-65$\mug, respectively. In this band, the \gls{ism} component plays a very minor, if not completely negligible, role because the low-energy particles involved in the radiation are released from the remnant very late.

In the low turbulence cases, most of the leptons injected into the nebula have escaped it past 1\kyr, and they are henceforth trapped in the \gls{snr}. The corresponding synchrotron emission is therefore about the same in all three model setups (for the three values of $\kappa_T$), and will hardly change from 1 to 10\kyr. The residual lepton population still residing in the \gls{pwn} depends significantly on the escape conditions (the values of $\kappa_T$), and so does the corresponding synchrotron emission. Despite this, the emission from the \gls{pwn} is initially stronger than that of the \gls{snr} because of a much stronger magnetic field in the latter. As time goes by, continued escape from the nebula and decrease of the magnetic field cause the radio signal from the \gls{pwn} to drop below that of the \gls{snr}. The situation is very similar in the high turbulence cases, except that the emission from the \gls{pwn} is weakly dependent on the turbulence scale. The main reason for both behaviours is that advection then dominates the escape, such that the details of turbulence matter little.

The model also predicts synchrotron emission in the X-ray range but we do not present or discuss it here as the one-zone assumption for the nebula seems little appropriate for this. X-ray emission from \glspl{pwn} is observed to be more concentrated around the pulsar than radio or gamma rays, in a region where the magnetic field is probably not representative of the average value in the nebula volume.

\end{document}